\documentclass[12pt]{article}
\usepackage{amssymb}
\usepackage{amsfonts}
\usepackage{amsmath}
\usepackage{graphicx}

\newtheorem{theorem}{Theorem}

\newtheorem{corollary}{Corollary}

\newtheorem{definition}{Definition}

\newtheorem{lemma}{Lemma}

\newtheorem{proposition}{Proposition}
\newtheorem{remark}{Remark}

\begin{document}

\title{Respecting Linear Orders for Supermajority Rules\thanks{%
Parts of the results in this manuscript are based on discussions with Minoru
Kitahara, to whom I would like to express my gratitude. An earlier version
of this paper was presented at the 2023 Spring Meeting of the Japanese
Economic Association, held at Kansai University. I would also like to
express my appreciation to Susumu Cato for his helpful comments and
suggestions. This work was supported by JSPS KAKENHI Grant Numbers 20K01675
and 22K01402.}}
\author{Yasunori Okumura\thanks{%
Address: 2-1-6, Etchujima, Koto-ku, Tokyo, 135-8533 Japan.
Phone:+81-3-5245-7300. Fax:+81-3-5245-7300. E-mail:
okuyasu@gs.econ.keio.ac.jp}}
\maketitle

\begin{center}
\textbf{Abstract}
\end{center}

We consider linear orders of finite alternatives constructed by aggregating
individual preferences. Specifically, we focus on linear orders that respect
modified collective preference relations derived from supermajority rules,
where modifications are introduced through two procedures if cycles occur.
One procedure utilizes the transitive closure, while the other employs the
Suzumura-consistent closure, ensuring the elimination of cycles through
consistency adjustments. We derive two sets of linear orders that respect
these modified collective preference relations derived from \textit{all}
supermajority rules, and show that these sets are generally nonempty. We
show that these sets of linear orders closely relate to those obtained by
the ranked pairs method and the Schulze method, thereby providing new
insights into these influential methods. Finally, we show that any linear
order belonging to either set satisfies two important properties: the
extended Condorcet criterion and the Pareto principle.


\newpage

\section{Introduction}

We consider linear orders of a finite set of alternatives obtained by
aggregating the preferences of multiple individuals. Although Arrow's (1963)
impossibility theorem applies to the construction of such linear orders,
several methods have been introduced in previous studies to derive linear
orders from individual preferences without requiring the independence of
irrelevant alternatives. These methods include the Borda method\footnote{%
Pattanaik {}(2002) calls that the method to have an order by using the Borda
count the Borda ranking rule. Here, we simply call it the Borda method.}
(Borda 1781), the Kemeny-Young method (Kemeny (1958), Young and Levenglick
(1978), Young (1988)), Tideman's ranked pairs method (Tideman (1987, 2006))
and the Schulze method (Schulze (2011, 2018)). A linear order is useful not
only for choosing one alternative but also for choosing $n(\geq 2)$
alternative(s) from $n+1$ or more, and for selecting backup alternatives
from among the options that have not been chosen.

We focus on sets of linear orders of alternatives derived as follows. First,
for a profile of the preferences of individuals and each of the
supermajority rules, collective (or social) preference relations are derived
by aggregating the preferences of individuals. Second, if the collective
preference relations obtained by the first step have some cycles \`{a} la
Condorcet (1784), then we modify it by using one of two procedures. Third,
for each supermajority rule, a set of linear orders that respect the
(modified) collective preference relations is derived. Finally, we derive
the intersection of the sets of these linear orders and show that this
intersection is nonempty even when the domain of individual preferences is
unrestricted. In summary, we derive the set of linear orders that respect
the collective preference relations generated by any supermajority rule,
with modifications if cycles are present.

We explain our approach and its advantages in more detail. To construct
collective preference relations, we utilize supermajority rules
characterized by a level parameter $\alpha \in \left[ 1/2,1\right) $. In the
supermajority rule with $\alpha \in \left[ 1/2,1\right) $, an alternative $a$
is (collectively) pairwisely preferred to alternative $b$ if and only if
more than $100\times \alpha \%$ of individuals prefer $a$ to $b$. Note that
if $\alpha =1/2$, then this is the simple majority rule; and if $\alpha $ is
almost $1$, then this is the Pareto rule. Let $R_{\alpha }$ denote the
collective preference relation generated by a given $\alpha \in \left[
1/2,1\right) $.

We say that a linear order that respects a preference relation if for any
two alternatives $a$ and $b$, whenever $a$ is strictly pairwise preferred to 
$b$, $a$ has a higher rank than $b$ in the linear order. If $R_{\alpha }$
has some cycles, then there is no linear order that respects $R_{\alpha }$
and, as shown by Greenberg (1979), $R_{\alpha }$ may have some cycles unless 
$\alpha $ is almost $1$ (i.e., the Pareto rule). Thus, we modify $R_{\alpha }
$ using two distinct procedures: the transitive closure procedure by Bordes
(1976)\footnote{%
According to the interpretation of Caplin and Nalebuff (1988), Condorcet
also considers the transitive closure procedure to immune to the cycles.}
and the Suzumura-consistent closure procedure.\footnote{%
Bossert et al. (2005) first define the Suzumura-consistency.} Denoting the
modified $R_{\alpha }$ by $T\left( R_{\alpha }\right) $ and $S\left(
R_{\alpha }\right) $ through the former and latter procedures respectively,
we derive sets of linear orders that respectively respect $T\left( R_{\alpha
}\right) $ and $S\left( R_{\alpha }\right) $ for all $\alpha \in \left[
1/2,1\right) $. Subsequently, we define the $T$\textbf{-order set} (resp.
the $S$\textbf{-order set})) as the intersection of sets of linear orders
that respect $T\left( R_{\alpha }\right) $ (resp. $S\left( R_{\alpha
}\right) $) for all $\alpha \in \left[ 1/2,1\right) $. We show that the $T$%
-order set is a subset of the $S$-order set and is nonempty for any profile
of individual preference relations.

An advantage of our approach is that it is not necessary to choose proper
levels of the supermajority rule, because we focus on the intersection of
the sets of linear orders that respectively respect $T\left( R_{\alpha
}\right) $ and $S\left( R_{\alpha }\right) $ for \textit{all} $\alpha \in %
\left[ 1/2,1\right) $. To the contrary, using a single level $\alpha $ would
cause us to overlook a significant amount of information. Specifically, when
we focus on $R_{\alpha }$, $\left( a,b\right) \notin R_{\alpha }$ (resp. $%
\left( a,b\right) \in R_{\alpha }$) even if just slightly fewer (resp. more
or exactly) than $100\times \alpha \%$ of individuals prefer $a$ to $b$.
That is, the differences in proportions are difficult to capture by relying
on a single value of one $\alpha $. On the other hand, our approach can
capture the differences in proportions, because we consider $R_{\alpha }$
for \textit{all} $\alpha \in \left[ 1/2,1\right) $.

Moreover, since adopting either a high or low value of $\alpha $ has both
advantages and disadvantages, as discussed later, determining an appropriate 
$\alpha $ is difficult. First, consider a low level $\alpha $. Related to
this case, Arrow (1977) presents the social choice method using $T\left(
R_{1/2}\right) $ as one of the promising methods.\footnote{%
Arrow (1977) mentions that this method is proposed by Donald Campbell and
Georges Bordes. Bordes (1976) introduce the characteristics of this method.}
However, a collective preference relation $R_{\alpha }$ with a low $\alpha $
typically contains many cycles. Thus, if $\alpha $ is low, then significant
modifications to the collective preference relation may be necessary. Then,
as warned by Arrow (1977), we may treat too many alternatives as indifferent
with the modifications. Suzumura (2012) highlights a more critical drawback
of this method; that is, in our terminology, even if every individual
prefers an alternative to another, the latter may have a higher rank than
that of the former in a linear order respecting $T\left( R_{1/2}\right) $.%
\footnote{%
Precisely speaking, Suzumura (2012) defines a social choice rule using the
transitive closure and provides a concrete example illustrating that this
rule fails to satisfy the Pareto principle.}

On the other hand, although cycles cannot be completely eliminated unless $%
\alpha $ is almost $1$, they become rare events if $\alpha $ is high. See
for example Caplin and Nalebuff (1988) and Balasko and Cr\`{e}s (1997) on
the fact.\footnote{%
A result of Nakamura (1979) also claims this fact, because the Nakamura
number, which is the largest number of alternatives ensuring no cycle (plus
one), is increasing in the number of individuals times $\alpha $.
\par
Further, there is a vast literature on the acyclic domain, which is the set
of preferences that guarantee nonexistence of cycles. On this literature,
Black (1958) is the pioneer and Greenberg (1979), Slutsuky (1979), Couglin
(1981), Caplin and Nalebuff (1988), Saari (2014) and Gjorgjiev and Xefteris
(2015), and Cr\`{e}s and \"{U}nver (2017) consider the acyclic domains with
supermajority rules. Their results imply that the domain would be wider when 
$\alpha $ rises.} Therefore, if $\alpha $ is high, then significant
modifications to the collective preference relation are not necessary.
However, as shown by Tovey (1997), the scarceness of cycles is a result of
the incompleteness of the collective preference relation.\footnote{%
Thus, Tovey (1997) states that \textquotedblleft An interesting research
problem would be how to modify the model to retain the scarceness of cycles
but permit more complete aggregate preferences.\textquotedblright\ In this
study, we try to solve this problem.} Thus, we may also treat too many
alternatives as indifferent if $\alpha $ is large. Moreover, the Condorcet
loser, which is the alternative that pairwisely loses any other with the
simple majority rule, may have the highest rank in a linear order respecting 
$T\left( R_{\alpha }\right) $ or $S\left( R_{\alpha }\right) $ with a
sufficiently large $\alpha $.

To understand these points, we consider the following example. Suppose that
there are five alternatives $a$, $b$, $c$, $d$ and $e$, where $e$ is an
inferior copy of $a$ and thus all individuals prefer $a$ to $e$. The ratio
of individuals who prefer $a$ and $e$ to $b$ is $0.7$, that of those who
prefer $b$ to $c$ is $0.66$ and that of those who prefer $c$ to $a$ and $e$
is $0.64$. Thus, there are Condorcet cycles. Moreover, the ratios of
individuals who prefer each of $a,b,c$ and $e$ to $d$ is $0.6$ and therefore 
$d$ is\ the Condorcet loser. In this case, the modified collective
preferences with the two procedures are the same for all $\alpha \in \left[
1/2,1\right) $. Then, the characteristics of the linear orders that respect
the modified collective preference relations with each $\alpha \in \left[
1/2,1\right) $ are given in the following table, where $x\succ y$ means that
the rank of $x$ must be higher than that of $y$ in any respecting linear
orders.

\begin{center}
\begin{tabular}{|l|l|}
\hline
$\alpha $ & Characteristics of respecting linear orders \\ \hline
$\left[ 0.5,0.6\right) $ & $d$ has the lowest rank \\ \hline
$\left[ 0.6,0.64\right) $ & Any \\ \hline
$\left[ 0.64,0.66\right) $ & $a\succ e,$ $a\succ b,$ $e\succ b,$ $b\succ c$
\\ \hline
$\left[ 0.66,0.7\right) $ & $a\succ e,$ $a\succ b,$ $e\succ b$ \\ \hline
$\left[ 0.7,1\right) $ & $a\succ e$ \\ \hline
\end{tabular}
\end{center}

If $0.64\leq \alpha <1$, then there is no cycle but $d$ can occupy the
highest position in some respecting linear orders. Therefore, when $\alpha $
is high, there may be too many respecting linear orders, including
inappropriate ones in which the Condorcet loser receives the highest rank.
On the other hand, if $0.5\leq \alpha <0.64$, then the position of $e$ is
higher than that of $a$ in some respecting linear orders. Thus, when $\alpha 
$ is low, there may also exist too many respecting linear orders, including
inappropriate ones in which one alternative is ranked higher than another
despite all individuals preferring the latter to the former.

This example implies that it is hard to decide one proper $\alpha \in \left[
1/2,1\right) $ for deriving a \textquotedblleft good\textquotedblright\
respecting linear order. However, in our approach, there is no need to
decide it at all. In fact, in the example, the linear order in the order of $%
a$, $e$, $b$, $c$ and $d$ is the unique linear order that is in the $T$%
-order set, which is in this case equivalent to the $S$-order set.

Our results are especially related to two influential voting methods: the
ranked pairs method (Tideman (1987, 2006)) and the Schulze method (Schulze
(2011, 2018)), which are two of the influential methods. Their common
feature is that they firstly derive a binary relation from the preferences
of individuals. As summarized by Tideman (2006), these two methods share
many important properties of voting rules. Moreover, the Schulze method is
used in several real world elections (Schulze (2018)). First, we show that
any linear order obtained by the ranked pairs method is in the $S$-order
set. Second, we show that the set of linear orders that can be obtained by
the Schulze method is equivalent to the $T$-order set. This result provides
novel characterization and interpretation of the Schulze method. Since the $%
T $-order set is a subset of the $S$-order set, any linear order obtained by
the Schulze method is also in the $S$-order set. These results imply both
the ranked pairs method and the Schulze method are in the class of methods
that choose one from the $S$-order set. Consequently, this clarifies that
the relationship between these methods, although not widely recognized, is
interconnected.

Finally, we show that any linear order belonging to the S-order set---which
includes the T-order set as a subset---satisfies two properties: the
extended Condorcet criterion and the Pareto principle, originally introduced
by Truchon (1998) and Arrow (1963), respectively.

\section{Model}

Let $A$ be the finite set of alternatives where $\left\vert A\right\vert
\geq 3$. Let $R\subseteq A\times A$ be a binary relation on $A$. In this
study, we focus only on a irreflexive binary relation; that is, $(a,a)\notin
R$ for all $a\in A$. The asymmetric part of $R$ denoted by $P\left( R\right) 
$ is a binary relation on $A$ satisfying $(a,b)\in P\left( R\right) $ if and
only if $(a,b)\in R$ and $(b,a)\notin R$. As mentioned particularly in the
subsequent section, a binary relation on $A$ denoted by $R$ represents not
only the preference of an individual but a collective (or social) preference
that is obtained by aggregating the preferences of a number of individuals
via each of the supermajority rules.

We define the following properties of binary relations. A binary relation $R$
on $A$ is

\begin{description}
\item \textbf{complete}\textit{\ }if for all $a,b\in A$ such that $a\neq b,$%
\textit{\ }$(a,b)\in R$ or $(b,a)\in R$,

\item \textbf{negatively} \textbf{transitive}\textit{\ }if for all $a,b,c\in
A,$ [$\left( a,b\right) \notin R$ and $\left( b,c\right) \notin R$] implies $%
\left( a,c\right) \notin R$,

\item \textbf{transitive} if for all $a,b,c\in A,$ [$(a,b)\in R$ and $%
(b,c)\in R$] implies $(a,c)\in R$,

\item \textbf{Suzumura} \textbf{consistent }if for all $K\in \mathbb{N}%
\setminus \left\{ 1\right\} $ and for all $a_{0},a_{1},\cdots ,a_{K}\in A$, $%
(a_{k-1},a_{k})\in R$ for all $k\in \left\{ 1,\cdots ,K\right\} $ implies $%
(a_{K},a_{0})\notin P\left( R\right) $,

\item \textbf{P-acyclic }if for all $K\in \mathbb{N}\setminus \left\{
1\right\} $ and for all $a_{0},a_{1},\cdots ,a_{K}\in A$, $%
(a_{k-1},a_{k})\in P\left( R\right) $ for all $k\in \left\{ 1,\cdots
,K\right\} $ implies $(a_{K},a_{0})\notin P\left( R\right) $.

\item \textbf{asymmetric} if for all $a,b\in A,$ $(a,b)\in R$ implies $%
(b,a)\notin R$.
\end{description}

Let a \textbf{partial order} be asymmetric and transitive binary relation.
Let a \textbf{weak order} be a partial order that satisfies negative
transitivity. Moreover, let a \textbf{linear order} be a weak order that
satisfies completeness. If $R$ is a linear order where $\left(
a_{1},a_{2}\right) ,\left( a_{2},a_{3}\right) ,\cdots ,\left( a_{\left\vert
A\right\vert -1},a_{\left\vert A\right\vert }\right) \in R$ and $%
a_{1},a_{2},\cdots ,a_{\left\vert A\right\vert }$ are distinct, then we
simply write 
\begin{equation*}
R:a_{1},a_{2},\cdots ,a_{\left\vert A\right\vert }\text{.}
\end{equation*}

A finite sequence of alternatives $\left( a_{1},a_{2},\cdots ,a_{m}\right) $
where $a_{i}\in A$ for all $i=1,\cdots ,m\left( \geq 2\right) $ is called a 
\textbf{path} \textbf{for} $R$ \textbf{from} $a$ \textbf{to} $b$ if $a_{1}=a$%
, $a_{m}=b$, $\left( a_{l},a_{l+1}\right) \in R$ for all $l=1,2,\cdots ,m-1$%
, and $a_{1},a_{2},\cdots ,a_{m}$ are distinct, where $m$ represents the
length of the path.\footnote{%
In several previous studies, a path\ for $R$ is referred to as a
\textquotedblleft beatpath\textquotedblright .} Let $\mathcal{B}\left( \left[
a,b\right] ,R\right) $ be the set all paths for $R$ from $a$ to $b$. If
there is a path for $R$ from $a$ to $b$ whose length is $2$; that is, if $%
\left( a,b\right) \in R$, then we say that $a$ \textbf{directly beats} $b$
for $R$. If the length of any path from $a$ to $b$ is $3$ or more, then we
say that $a$ \textbf{indirectly beats} $b$, for $R$. If there are two paths
for $R$ from $a$ to $b$ and that from $b$ to $a$, then there is a \textbf{%
cycle }for $R$ (containing $a$ and $b$). A cycle for $P\left( R\right) $ is
called a \textbf{P-cycle} for $R$. Trivially, if $R$ is P-acyclic, then
there is no P-cycle for $R$.

\subsection{Collective preference relations}

Let $I$ be the set of finite individuals (or voters) with $\left\vert
I\right\vert \geq 3$. Each individual $i\in I$ has a binary relation $R^{i}$
over $A$ where $\left( a,b\right) \in R^{i}$ (resp. $\left( a,b\right) \in
P\left( R^{i}\right) $) for $a,b\in A$ implies that individual $i$ weakly
(resp. strictly) prefers $a$ to $b$. We allow that $R^{i}$ is not a weak
order for some $i\in I$; that is, we can examine a general situation
considered by several recent studies such as Ohbo et al. (2005), Barber\`{a}
and Bossert (2023), Fujiwara-Greve et al. (2023), Villar (2023) and Okumura
(2024).

We introduce the supermajority rules for constructing collective preference
relations by aggregating the preferences of individuals. Let 
\begin{equation*}
N\left[ a,b\right] =\left\vert \left\{ i\in I\text{ }\left\vert \text{ }%
\left( a,b\right) \in P\left( R^{i}\right) \right. \right\} \right\vert ,
\end{equation*}%
which represents the number of individuals who strictly prefer $a$ to $b$.
Thus, if $\left( a,b\right) \in R^{i}$ and $\left( b,a\right) \in R^{i}$,
then neither is counted. Note that $N\left[ a,b\right] +N\left[ b,a\right]
\leq \left\vert I\right\vert $.

Moreover, let $f:\mathbb{Z}_{+}\times \mathbb{Z}_{+}\rightarrow \left[ 0,1%
\right] $ be such that for any $n\geq 0$ and $m\geq 0$ satisfying $n+m\leq
\left\vert I\right\vert $, $f\left( n+1,m\right) >f\left( n,m\right) $, $%
f\left( n,m+1\right) <f\left( n,m\right) $, and $f\left( n,m\right) =1/2$ if 
$n=m$. For example, $f\left( n,m\right) $ can be%
\begin{eqnarray}
f\left( n,m\right) &=&\frac{n}{n+m}\text{ if }n+m>0  \label{b} \\
&=&\frac{1}{2}\text{ if }n+m=0,\text{ or}  \notag
\end{eqnarray}%
\begin{equation}
f\left( n,m\right) =\frac{\left\vert I\right\vert +\left( n-m\right) }{%
2\left\vert I\right\vert }\text{.}  \label{b1}
\end{equation}

For notational simplicity, we let $\Phi \left[ a,b\right] =f\left( N\left[
a,b\right] ,N\left[ b,a\right] \right) $. If $R^{i}$ is a linear order for
all $i\in I$, then $\Phi \left[ \cdot ,\cdot \right] $ defined by (\ref{b})
is equivalent to that defined by (\ref{b1}), because, in that case, $N\left[
a,b\right] +N\left[ b,a\right] =\left\vert I\right\vert $.

We let $R_{\alpha }\subseteq A\times A$ be a collective preference relation
for $\alpha \in \left[ 1/2,1\right) $ that satisfies $\left( a,b\right) \in
R_{\alpha }$ if and only if $\Phi \left[ a,b\right] >\alpha $. In the
terminology of Schulze (2011), $\Phi \left[ a,b\right] $ represents the
degree of strength of the link between $a$ and $b$. Moreover, if (\ref{b})
and (\ref{b1}) are used, then the strength of a link is measured by the
ratio and the margin, respectively.

We consider an example to understand the construction of the binary
relations.

\subsubsection*{Example 1}

There are $A=\left\{ a,b,c,d,e\right\} $ and $I=\left\{ 1,\cdots ,9\right\} $%
. All individuals have a linear preference order on $A$ and%
\begin{gather*}
R^{i}:\text{ }d,a,e,b,c,\text{ for all }i=1,2,3, \\
R^{i^{\prime }}:\text{ }b,c,a,e,d,\text{ for all }i^{\prime }=4,5,6, \\
R^{i^{\prime \prime }}:\text{ }c,a,e,b,d,\text{ for all }i^{\prime \prime
}=7,8, \\
R^{9}:\text{ }d,c,a,e,b.
\end{gather*}%
Since this is a linear order, $N\left[ x,y\right] =9-N\left[ y,x\right] $
for all $x,y\in A$ such that $x\neq y$. Since $N\left[ x,d\right] =5$ for
all $x\in A\setminus \left\{ d\right\} $, $d$ is the Condorcet loser.
Moreover, $N\left[ a,e\right] =9$; that is, all individuals prefer $a$ to $e$%
. Moreover, 
\begin{equation*}
N\left[ a,b\right] =N\left[ e,b\right] =N\left[ b,c\right] =N\left[ c,a%
\right] =N\left[ c,e\right] =6
\end{equation*}%
When we adopt (\ref{b}) or (\ref{b1}), 
\begin{equation*}
\Phi \left[ a,b\right] =\Phi \left[ e,b\right] =\Phi \left[ b,c\right] =\Phi %
\left[ c,a\right] =\Phi \left[ c,e\right] =2/3
\end{equation*}%
and $\Phi \left[ x,d\right] =5/9$ for all $x\in A\setminus \left\{ d\right\} 
$. Let $\alpha _{1}\in \left[ 1/2,5/9\right) ,$ $\alpha _{2}\in \left[
5/9,2/3\right) $ and $\alpha _{3}\in \left[ 2/3,1\right) $. Then,%
\begin{eqnarray*}
R_{\alpha _{1}} &=&\left\{ \left( a,b\right) ,(e,b),\left( b,c\right)
,\left( c,a\right) ,(c,e),\left( a,e\right) ,\left( a,d\right) ,\left(
b,d\right) ,\left( c,d\right) ,\left( e,d\right) \right\} \\
R_{\alpha _{2}} &=&\left\{ \left( a,b\right) ,(e,b),\left( b,c\right)
,\left( c,a\right) ,(c,e),(a,e)\right\} \\
R_{\alpha _{3}} &=&\left\{ \left( a,e\right) \right\} .
\end{eqnarray*}

Hereafter, we arbitrarily fix $f$ and a profile of the preference relations
of all individuals $\left( R^{i}\right) _{i\in I}$.$\ $Thus, $R_{\alpha }$
for all $\alpha \in \left[ 1/2,1\right) $ are given.

\subsection{Respecting Linear Orders}

We introduce the linear orders that respect a binary relation $R$ on $A$.
The formal definition is as follows.

\begin{definition}
For a binary relation $R$ on $A$, let $\mathcal{R}\left( R\right) $ be the
set of linear orders that \textbf{respect }$R$ if any $\hat{R}\in \mathcal{R}%
\left( R\right) $ is a linear order on $A$ and $\left( a,b\right) \in
P\left( R\right) $ implies $\left( a,b\right) \in \hat{R}$.
\end{definition}

That is, for any $\hat{R}\in \mathcal{R}\left( R\right) $, if $\left(
a,b\right) \in P\left( R\right) $, then $\left( a,b\right) \in \hat{R}$ and
since $\hat{R}$ is a linear order, $\left( a,b\right) \in P\left( \hat{R}%
\right) $. On the other hand, if either [$\left( a,b\right) \in R$ and $%
\left( b,a\right) \in R$] or [$\left( a,b\right) \notin R$ and $\left(
b,a\right) \notin R$] holds, either $\left( a,b\right) \in \hat{R}$ or $%
\left( b,a\right) \in \hat{R}$ is allowed to hold.

In the terminology of set theory, $\hat{R}\in \mathcal{R}\left( R\right) $
if and only if $\hat{R}$ is a linear order extension of $P\left( R\right) $;
that is, $\hat{R}$ is a superset of $P\left( R\right) $ and a linear order.
Thus, if $R$ is asymmetric, $\hat{R}\in \mathcal{R}\left( R\right) $ if and
only if $\hat{R}$ is a linear order extension of $R$. However, we allow that 
$R$ is not asymmetric and in that case, $\hat{R}\in \mathcal{R}\left(
R\right) $ is not necessarily a linear order extension of $R$, although it
remains a linear order extension of $P\left( R\right) $.

We provide a class of algorithms to construct a linear order $\hat{R}\in 
\mathcal{R}\left( R\right) $ for a P-acyclic $R$.

First, for $A^{\prime }\subseteq A$ a binary relation $R$, let $M\left(
A^{\prime },R\right) $ be a maximal set; that is, 
\begin{equation*}
M\left( A^{\prime },R\right) =\left\{ a\in A^{\prime }\text{ }\left\vert 
\text{ }\left( b,a\right) \notin P\left( R\right) \text{ for all }b\in
A^{\prime }\setminus \left\{ a\right\} \right. \right\} .
\end{equation*}

We consider the following algorithms.\newline

\textbf{Step} $1$: Let $a_{1}\in M\left( A,R\right) $.

\textbf{Step} $t=2,\cdots ,\left\vert A\right\vert $: Let $a_{t}\in M\left(
A\setminus \left\{ a_{1},\cdots ,a_{t-1}\right\} ,R\right) $.\newline

Finally, we have a linear order $\hat{R}:a_{1},\cdots ,a_{\left\vert
A\right\vert }$.

Since $M\left( A^{\prime },R\right) $ may not be a singleton for some $%
A^{\prime }\subseteq A$, this is a class of algorithm that we call the 
\textbf{sequential maximal ordering} (hereafter \textbf{SMO}) class. If $R$
is P-acyclic, then each algorithm in the SMO class results in a linear order
that respects $R$, because as shown by Bossert and Suzumura (2010), $R$ is
P-acyclic and thus the maximal set is not empty for any step.

\begin{proposition}
$\mathcal{R}\left( R\right) \neq \emptyset $ if and only if $R$ is
P-acyclic. Moreover, a linear order respects $R$ if and only if it is
obtained by an algorithm within the SMO class with $R$.
\end{proposition}

Since $R_{\alpha }$ introduced above may not be P-acyclic unless $\alpha $
is almost 1, by Proposition 1, there may be no linear order that respects $%
R_{\alpha }$. Thus, we modify $R_{\alpha }$ to have its respecting linear
order, even if $R_{\alpha }$ has some P-cycle.

\subsection{Transitive and Suzumura-consistent Closures}

First, a binary relation $T\left( R\right) $ is a \textbf{transitive closure}
of $R$ if $T\left( R\right) $ is the smallest transitive binary relation
containing $R$. This modification is introduced by Bordes (1976) to escape
from Condorcet cycles.\footnote{%
Moreover, note that as is shown by Deb (1977), $M\left( A,T\left( R\right)
\right) $ is equivalent to the Schwartz set for an asymmetric $R$.} By
Bossert and Suzumura (2010, Theorem 2.1), $T\left( R\right) $ is calculated
as, for a binary relation $R$, $\left( a,b\right) \in T\left( R\right) $ if
and only if $\mathcal{B}\left( \left[ a,b\right] ,R\right) \neq \emptyset $.

Second, a binary relation $S\left( R\right) $ is a \textbf{%
Suzumura-consistent closure} of $R$ if $S\left( R\right) $ is the smallest
Suzumura-consistent binary relation containing $R$. Bossert and Suzumura
(2010, Theorem 2.8) show that for a binary relation $R$,%
\begin{equation*}
S\left( R\right) =R\cup \left\{ \left. \left( a,b\right) \in A\times A\text{ 
}\right\vert \text{ }\left( a,b\right) \in T\left( R\right) \text{ and }%
\left( b,a\right) \in R\right\} .
\end{equation*}

To see how the modification is done, we provide an example.

\subsubsection*{Example 2}

Let $A=\left\{ a,b,c,d\right\} $ and $I=\{1,\cdots ,8\}$. Moreover,%
\begin{gather*}
R^{1}:\text{ }a,b,c,d,\text{ }R^{2}:\text{ }a,b,d,c,\,R^{3}:\text{ }d,c,a,b,
\\
R^{4}:\text{ }c,d,a,b,\text{ }R^{5}:\text{ }b,c,a,d,\,R^{6}:\text{ }b,d,c,a,
\\
R^{7}:\text{ }c,b,a,d,\text{ }R^{8}:\text{ }a,d,b,c.
\end{gather*}%
Then, 
\begin{eqnarray*}
N\left[ a,b\right]  &=&N\left[ b,c\right] =N\left[ c,a\right] =N\left[ a,d%
\right] =N\left[ b,d\right] =5 \\
N\left[ c,d\right]  &=&N\left[ d,c\right] =4.
\end{eqnarray*}%
Fix $\alpha \in \left[ 1/2,5/8\right) $. Then,%
\begin{equation*}
R_{\alpha }=\left\{ \left( a,b\right) ,\left( b,c\right) ,\left( c,a\right)
,\left( a,d\right) ,\left( b,d\right) \right\} ,
\end{equation*}%
and there is a cycle of $R_{\alpha }$ ($a\rightarrow b\rightarrow
c\rightarrow a$). By Proposition 1, there is no linear order that respects $%
R_{\alpha }$. We derive $T\left( R_{\alpha }\right) $ and $S\left( R_{\alpha
}\right) $. Then,

\begin{eqnarray*}
T\left( R_{\alpha }\right) &=&R_{\alpha }\cup \left\{ \left( b,a\right)
,\left( c,b\right) ,\left( a,c\right) ,\left( c,d\right) \right\} , \\
S\left( R_{\alpha }\right) &=&R_{\alpha }\cup \left\{ \left( b,a\right)
,\left( c,b\right) ,\left( a,c\right) \right\} .
\end{eqnarray*}%
Note that $\left( c,a\right) ,\left( a,d\right) \in R_{\alpha }$ implies $%
\left( c,d\right) \in T\left( R_{\alpha }\right) $, but $\left( d,c\right)
\notin R_{\alpha }$ implies $\left( d,c\right) \notin S\left( R_{\alpha
}\right) $.

Any linear orders $R^{\prime }$ satisfying $\left( a,d\right) ,\left(
b,d\right) \in R^{\prime }$ are included in $\mathcal{R}\left( S\left(
R_{\alpha }\right) \right) \left( \supseteq \mathcal{R}\left( T\left(
R_{\alpha }\right) \right) \right) $. However, since $\left( c,d\right) \in
T\left( R_{\alpha }\right) $, $\mathcal{R}\left( T\left( R_{\alpha }\right)
\right) $ only includes the linear orders satisfying that $d$ has the lowest
rank. For example, if $R^{\prime }:b,a,d,c$, then $R^{\prime }\in \mathcal{R}%
\left( S\left( R_{\alpha }\right) \right) $ but $R^{\prime }\notin \mathcal{R%
}\left( T\left( R_{\alpha }\right) \right) $. Thus, $\mathcal{R}\left(
T\left( R_{\alpha }\right) \right) \subsetneq \mathcal{R}\left( S\left(
R_{\alpha }\right) \right) $ in this example.\newline

Now, we arbitrarily fix $\left( R^{i}\right) _{i\in I}$ and $\alpha \in %
\left[ 1/2,1\right) $ and consider $R_{\alpha }$. Recall that $R_{\alpha }$
may have a P-cycle. We have the following result that ensures that both $%
T\left( R_{\alpha }\right) $ and $S\left( R_{\alpha }\right) $ can be used
as suitable modifications of $R_{\alpha }$.

\begin{proposition}
First, 
\begin{equation*}
\emptyset \neq \mathcal{R}\left( T\left( R_{\alpha }\right) \right)
\subseteq \mathcal{R}\left( S\left( R_{\alpha }\right) \right) \text{,}
\end{equation*}%
for all $\alpha \in \left[ 1/2,1\right) $. Second, if $\mathcal{R}\left(
R_{\alpha }\right) \neq \emptyset $, then 
\begin{equation*}
\mathcal{R}\left( T\left( R_{\alpha }\right) \right) =\mathcal{R}\left(
S\left( R_{\alpha }\right) \right) =\mathcal{R}\left( R_{\alpha }\right) 
\text{,}
\end{equation*}%
for all $\alpha \in \left[ 1/2,1\right) $.
\end{proposition}

By the first result, both $\mathcal{R}\left( T\left( R_{\alpha }\right)
\right) $ and $\mathcal{R}\left( S\left( R_{\alpha }\right) \right) $ are
nonempty, because $T\left( R_{\alpha }\right) $ and $S\left( R_{\alpha
}\right) $ are P-acyclic (see Lemma 1 provided in Appendix A). Moreover, by
the second result, the modification is done only if $\mathcal{R}\left(
R_{\alpha }\right) =\emptyset $.

We immediately have $R_{\alpha }\subseteq S\left( R_{\alpha }\right)
\subseteq T\left( R_{\alpha }\right) $ and thus $S\left( R_{\alpha }\right) $
is a smaller modification of $R_{\alpha }$ comparing to $T\left( R_{\alpha
}\right) $. However, $\mathcal{R}\left( T\left( R_{\alpha }\right) \right)
\subseteq \mathcal{R}\left( S\left( R_{\alpha }\right) \right) $ implies
that the set of linear orders respecting $T\left( R_{\alpha }\right) $ is a
refinement of that with $S\left( R_{\alpha }\right) $.

\subsection{Respecting linear orders with supermajority rule}

Here, we consider the linear orders respecting the modified $R_{\alpha }$,
which is either $S\left( R_{\alpha }\right) $ or $T\left( R_{\alpha }\right) 
$, for specific $\alpha \in \left[ 1/2,1\right) $. We show that both $%
\mathcal{R}\left( T\left( R_{\alpha }\right) \right) $ and $\mathcal{R}%
\left( S\left( R_{\alpha }\right) \right) $ may include some improper linear
orders for any $\alpha \in \left[ 1/2,1\right) $. To show this fact, we
revisit Example 1.

First, consider $\alpha _{2}\in \left[ 5/9,2/3\right) $. Then, since 
\begin{equation*}
P\left( T\left( R_{\alpha _{2}}\right) \right) =P\left( S\left( R_{\alpha
_{2}}\right) \right) =\emptyset ,
\end{equation*}%
any linear orders of $\left\{ a,b,c,d,e\right\} $ including $\bar{R}:$ $%
d,a,b,c,e$ and $\bar{R}^{\prime }:$ $e,a,b,c,d$ are in $\mathcal{R}\left(
T\left( R_{\alpha _{2}}\right) \right) \left( =\mathcal{R}\left( S\left(
R_{\alpha _{2}}\right) \right) \right) $. However, first, the Condorcet
loser $d$ is placed in the first position in $\bar{R}$. Second, although all
individuals unanimously prefer $a$ to $e$, $e$ ranks higher than $a$ in $%
\bar{R}^{\prime }$. Therefore, in this case, $\mathcal{R}\left( T\left(
R_{\alpha _{2}}\right) \right) $ includes some improper linear orders.

Second, consider $\alpha _{1}\in \left[ 1/2,5/9\right) $. Then,%
\begin{gather*}
T\left( R_{\alpha _{1}}\right) =S\left( R_{\alpha _{1}}\right) = \\
R_{\alpha _{1}}\cup \left\{ \left( b,a\right) ,\left( b,e\right) ,\left(
c,b\right) ,\left( b,a\right) ,\left( a,c\right) ,\left( e,c\right)
,(e,a)\right\} .
\end{gather*}%
Therefore, 
\begin{equation*}
P\left( T\left( R_{\alpha _{1}}\right) \right) =P\left( S\left( R_{\alpha
_{1}}\right) \right) =\left\{ \left( a,d\right) ,\left( b,d\right) ,\left(
c,d\right) ,\left( e,d\right) \right\} .
\end{equation*}%
Hence any linear orders such that $d$ is placed in the last position are
included in $\mathcal{R}\left( T\left( R_{\alpha _{1}}\right) \right) $.
Thus, $\bar{R}\notin \mathcal{R}\left( T\left( R_{\alpha _{1}}\right)
\right) $ but $\bar{R}^{\prime }\in \mathcal{R}\left( T\left( R_{\alpha
_{1}}\right) \right) $ implying that $\mathcal{R}\left( T\left( R_{\alpha
_{1}}\right) \right) $ also includes some improper linear orders. A similar
observation is presented by Suzumura (2012).

Finally, consider $\alpha _{3}\in \left[ 2/3,1\right) $. Then, 
\begin{equation*}
R_{\alpha _{3}}=T\left( R_{\alpha _{3}}\right) =S\left( R_{\alpha
_{3}}\right) =\left\{ \left( a,e\right) \right\} =P\left( R_{\alpha
_{3}}\right)
\end{equation*}%
has no P-cycle. Therefore, any linear orders such that $a$ ranks higher than 
$e$ are included in $\mathcal{R}\left( R_{\alpha _{3}}\right) $ ($=\mathcal{R%
}\left( T\left( R_{\alpha _{3}}\right) \right) =\mathcal{R}\left( S\left(
R_{\alpha _{3}}\right) \right) $ by Proposition 2). Thus, $\bar{R}^{\prime
}\notin \mathcal{R}\left( T\left( R_{\alpha _{3}}\right) \right) $ but $\bar{%
R}\in \mathcal{R}\left( T\left( R_{\alpha _{3}}\right) \right) $ implying
that $\mathcal{R}\left( T\left( R_{\alpha _{3}}\right) \right) $ also
includes some improper linear orders.

Thus, $\mathcal{R}\left( T\left( R_{\alpha }\right) \right) $ and $\mathcal{R%
}\left( S\left( R_{\alpha }\right) \right) $ may include some improper
linear orders for any $\alpha \in \left[ 1/2,1\right) $.

\section{Order sets}

First, we focus only on P-acyclic $R_{\alpha }$. Let $\mathcal{P}_{^{AC}}%
\mathcal{=}$ $\left\{ \left. R_{\alpha }\text{ }\right\vert \text{ }%
R_{\alpha }\text{ is P-acyclic}\right\} $. Note that if $\alpha $ is
sufficiently large (the Pareto rule) and $R^{i}$ is a weak order for all $%
i\in I,$ then $R_{\alpha }\in \mathcal{P}_{^{AC}}$ (see Lemma 9 provided in
Appendix A). Thus, in that case, we can let 
\begin{equation*}
\alpha ^{\ast }=\min_{\alpha \in \left[ 1/2,1\right) }\left\{ \left. \alpha 
\text{ }\right\vert \text{ }R_{\alpha }\in \mathcal{P}_{^{AC}}\right\} .
\end{equation*}%
On the other hand, if $R^{i}$ is not a weak order but a partial order\ for
some $i\in I$, then $\mathcal{P}_{AC}$ may be empty and thus $\alpha ^{\ast
} $ may not be well-defined. For example, if $I=\{1,2,3\},$ $A=\left\{
a,b,c\right\} ,$ $R^{1}=\left\{ (a,b)\right\} ,$ $R^{2}=\left\{
(b,c)\right\} $ and $R^{3}=\left\{ (c,a)\right\} ,$ and (\ref{b}) is used,
then $\mathcal{P}_{AC}$ is nonempty.

We have the following result.

\begin{proposition}
If $R^{i}$ is a weak order for all $i\in I$,%
\begin{equation*}
\bigcap\limits_{\alpha :\text{ }R_{\alpha }\in \mathcal{P}_{^{AC}}}\mathcal{R%
}\left( R_{\alpha }\right) =\mathcal{R}\left( R_{\alpha ^{\ast }}\right)
\neq \emptyset .
\end{equation*}
\end{proposition}

Considering P-acyclic $R_{\alpha }$ only is in a similar spirit to the
Simpson-Kramer min-max method introduced in the last section.\footnote{%
However, for any $\hat{R}\left( :a_{1},\cdots ,a_{\left\vert I\right\vert
}\right) \in \bigcap\nolimits_{\alpha :\text{ }R_{\alpha }\in \mathcal{P}%
_{^{AC}}}\mathcal{C}\left( R_{\alpha }\right) $, $a_{1}$ may not be the
Simpson-Kramer min-max winner. The Simpson-Kramer min-max method is formally
introduced in the last section.} However, the set $\bigcap\nolimits_{\alpha :%
\text{ }R_{\alpha }\in \mathcal{P}_{^{AC}}}\mathcal{R}\left( R_{\alpha
}\right) $ may include some improper linear orders as discussed above. To
illustrate this fact, we reconsider Example 1.

From the analysis above, we can see that in Example 1, $\alpha ^{\ast }=2/3$
and $R_{\alpha _{3}}=\left\{ \left( a,e\right) \right\} $ for all $\alpha
_{3}\geq \alpha ^{\ast }=2/3$. Then, $\bigcap\nolimits_{\alpha :\text{ }%
R_{\alpha }\in \mathcal{P}_{^{AC}}}\mathcal{R}\left( R_{\alpha }\right) =%
\mathcal{R}\left( T\left( R_{\alpha ^{\ast }}\right) \right) $ includes $%
\bar{R}$, which is an improper linear order, because where the Condorcet
loser is placed in the first position in $\bar{R}$. This occurs because, in
this approach, to avoid cycles, we disregard the fact that five out of nine
individuals (which is less than the required two-thirds threshold) rank
alternative $d$ lower than all other alternatives. Hence, $%
\bigcap\nolimits_{\alpha :\text{ }R_{\alpha }\in \mathcal{P}_{^{AC}}}%
\mathcal{R}\left( R_{\alpha }\right) $ may include some improper linear
orders. Thus, we consider a smaller set of linear orders that does not
include such improper linear orders.

We consider a procedure to use the Suzumura-consistent closure of $R_{\alpha
}$. By Proposition 2, even if $\mathcal{R}\left( R_{\alpha }\right)
=\emptyset $, $S\left( R_{\alpha }\right) $ must be nonempty. Therefore, we
can consider a linear order that respects $S\left( R_{\alpha }\right) $ for
all $\alpha \in \left[ 1/2,1\right) $. However, as shown above, $\mathcal{R}%
\left( S\left( R_{\alpha }\right) \right) $ may include some improper linear
orders for any $\alpha \in \left[ 1/2,1\right) $. Thus, we consider the $S$%
\textbf{-order set} defined as 
\begin{equation}
\mathcal{S}=\bigcap\limits_{\alpha \in \left[ 1/2,1\right) }\mathcal{R}%
\left( S\left( R_{\alpha }\right) \right) ,  \label{s}
\end{equation}%
which includes only linear orders that simultaneously respect $S\left(
R_{\alpha }\right) $ for \textit{all} $\alpha \in \left[ 1/2,1\right) $. We
have the following result.

\begin{theorem}
\begin{equation}
\mathcal{S}=\mathcal{R}\left( \bigcup_{\alpha \in \left[ 1/2,1\right)
}P\left( S\left( R_{\alpha }\right) \right) \right)  \label{e}
\end{equation}
\end{theorem}

To better understand Theorem 1, we reconsider Example 1.\ As explained
above, $\mathcal{R}\left( S\left( R_{\alpha _{1}}\right) \right) $ includes
the linear orders such that $d$ is placed in the last position, $\mathcal{R}%
\left( S\left( R_{\alpha _{2}}\right) \right) $ include all possible linear
orders, and $\mathcal{R}\left( S\left( R_{\alpha _{3}}\right) \right) $
include the linear orders such that $a$ ranks higher than $e$. Therefore, in
this example, the $S$-order set is equivalent to the set of the linear
orders such that $d$ is placed in the last position and $a$ ranks higher
than $e$.

We focus on $\bigcup_{\alpha \in \left[ 1/2,1\right) }P\left( S\left(
R_{\alpha }\right) \right) $. In this case, since 
\begin{eqnarray*}
P\left( S\left( R_{\alpha _{1}}\right) \right) &=&\left\{ \left( a,d\right)
,\left( b,d\right) ,\left( c,d\right) ,\left( e,d\right) \right\} , \\
P\left( S\left( R_{\alpha _{2}}\right) \right) &=&\emptyset ,\text{ } \\
P\left( S\left( R_{\alpha _{3}}\right) \right) &=&\left\{ \left( a,e\right)
\right\} ,\text{ and thus,} \\
\bigcup_{\alpha \in \left[ 1/2,1\right) }P\left( S\left( R_{\alpha }\right)
\right) &=&\left\{ \left( a,d\right) ,\left( b,d\right) ,\left( c,d\right)
,\left( e,d\right) ,\left( a,e\right) \right\} .
\end{eqnarray*}%
This implies that $\mathcal{R}\left( \bigcup_{\alpha \in \left[ 1/2,1\right)
}P\left( S\left( R_{\alpha }\right) \right) \right) $ must be the set of the
linear orders such that $d$ is placed in the last position and $a$ ranks
higher than $e$. Hence we can confirm that (\ref{e}) is satisfied in this
example.

Theorem 1 is useful to show the nonemptyness of $\mathcal{S}$ and to
introduce a comutationally efficient method to derive it. We present the
method in the subsequent section. To show the nonemptyness of it, we
consider the \textbf{ranked pairs method,} which is a rule to construct a
linear order from $\left( R^{i}\right) _{i\in I}$ and introduced by Tideman
(1987).

Let $\pi :A\times A\rightarrow \left\{ 1,2,\cdots ,\left\vert A\right\vert
^{2}\right\} $ be such that $\pi \left( a,b\right) \neq \pi \left(
c,d\right) $ for any $\left( a,b\right) \neq \left( c,d\right) $ and, if $%
\Phi \left[ a,b\right] >\Phi \lbrack c,d]$, then $\pi \left( a,b\right) <\pi
\left( c,d\right) $. Let $\Pi $ be a set of all possible $\pi $.

\textbf{Step 0 }Fix $\pi \in \Pi $.

\textbf{Step }$z\left( =1,\cdots ,\left\vert A\right\vert ^{2}\right) $%
\textbf{\ }Let $\left( a,b\right) \in A\times A$ be such that $\pi \left(
a,b\right) =z$. If $a=b$ or $\mathcal{B}\left( \left[ b,a\right]
,R_{z-1}^{\pi }\right) \neq \emptyset $, then let $R_{z}^{\pi }=R_{z-1}^{\pi
}$. Otherwise, then let $R_{z}^{\pi }=R_{z-1}^{\pi }\cup \left\{ \left(
a,b\right) \right\} $.

For the result of this algorithm, we let $R_{\left\vert A\right\vert
^{2}}^{\pi }=R^{\pi }$.

\begin{proposition}
For any $\pi \in \Pi $,%
\begin{equation}
R^{\pi }\in \mathcal{R}\left( \bigcup_{\alpha \in \left[ 1/2,1\right) \
}P\left( S\left( R_{\alpha }\right) \right) \right) =\mathcal{S}.  \label{f}
\end{equation}
\end{proposition}

This result implies that any linear order obtained by the ranked pairs
method respects the Suzumura-consistent closures of all supermajority binary
relations. Consequently, we obtain the following result.

\begin{corollary}
$\mathcal{S}\neq \emptyset $ and $\bigcup_{\alpha \in \left[ 1/2,1\right) \
}P\left( S\left( R_{\alpha }\right) \right) $ is P-acyclic.
\end{corollary}

Next, we consider a method to use the transitive closure of $R$. Let 
\begin{equation*}
\mathcal{T}=\bigcap\limits_{\alpha \in \left[ 1/2,1\right) }\mathcal{R}%
\left( T\left( R_{\alpha }\right) \right)
\end{equation*}%
be the $T$\textbf{-order set}.

First, we show that $\mathcal{T}$ has a similar characteristic to $\mathcal{S%
}$ as introduced in Theorem 1.

\begin{theorem}
\begin{equation*}
\mathcal{T}=\mathcal{R}\left( \bigcup_{\alpha \in \left[ 1/2,1\right)
}P\left( T\left( R_{\alpha }\right) \right) \right)
\end{equation*}
\end{theorem}

Since we can show this result by following the proof of Theorem 2, we omit
it.

Next, we immediately have the following result.

\begin{proposition}
First, $\mathcal{T\subseteq S}$. Second, if $R^{i}$ is a weak order for all $%
i\in I$, then%
\begin{equation*}
\mathcal{T\subseteq S}\subseteq \bigcap\limits_{\alpha :\text{ }R_{\alpha
}\in \mathcal{P}_{AC}}\mathcal{R}\left( R_{\alpha }\right) .
\end{equation*}
\end{proposition}

Note that $\mathcal{T\subsetneq S}$ may be satisfied. To show this, we
revisit Example 2. In this example, $\mathcal{S}$ is the set of linear
orders including $\left( b,d\right) $ and $\left( c,d\right) $. On the other
hand, $\mathcal{T\ }$is the set of linear orders in which $d$ has the lowest
rank. Therefore, in this case, $\mathcal{T\subsetneq S}$.

Therefore, when $\mathcal{T}$ is nonempty, $\mathcal{T}$ and $\mathcal{S}$
are refinements of $\bigcap\nolimits_{\alpha :\text{ }R_{\alpha }\in 
\mathcal{P}_{AC}}\mathcal{R}\left( R_{\alpha }\right) $ if $R^{i}$ is a weak
order for all $i\in I$, and $\mathcal{T}$ is a refinement of $\mathcal{S}$.
As shown in Appendix C, any linear order derived by the ranked pairs method
may not be an element of $\mathcal{T}$. Thus, to show the nonemptyness of $%
\mathcal{T}$, we consider another method.

For $\left( a_{1},a_{2},\cdots ,a_{m}\right) \in \mathcal{B}\left[ \left(
a,b\right) ,R_{1/2}\right] ,$ let the \textbf{strength} of $\left(
a_{1},a_{2},\cdots ,a_{m}\right) $ be 
\begin{equation*}
ST\left( \left( a_{1},\cdots ,a_{m}\right) \right) =\min_{j=1,\cdots ,m-1} 
\left[ \Phi \left[ a_{j},a_{j+1}\right] \right] .
\end{equation*}%
Furthermore, let 
\begin{align*}
B[a,b]& =\max_{p\in \mathcal{B}\left[ \left( a,b\right) ,R_{1/2}\right]
}ST\left( p\right) \text{ if}\mathrm{\ }\mathcal{B}\left[ \left( a,b\right)
,R_{1/2}\right] \neq \emptyset \\
& =0\text{ if\textrm{\ }}\mathcal{B}\left[ \left( a,b\right) ,R_{1/2}\right]
=\emptyset ,
\end{align*}%
which represents the strength of the strongest path from $a$ to $b$. Let $%
R^{Sc}$ be a binary relation obtained by the \textbf{Schulze method}; that
is, $\left( a,b\right) \in R^{Sc}$ if and only if $B[a,b]>B[b,a]$. Note
that, as shown by Schulze (2011), $R^{Sc}$ is a partial order and thus $%
\mathcal{R}\left( R^{Sc}\right) \neq \emptyset $ (see also Lemma 1 in
Appendix A and Proposition 1 above).

We have the following result.

\begin{proposition}
\begin{equation*}
R^{Sc}=\bigcup_{\alpha \in \left[ 1/2,1\right) }P\left( T\left( R_{\alpha
}\right) \right) .
\end{equation*}
\end{proposition}

This result provides a description of the Schulze method; that is, the
partial order $R^{Sc}$ derived by the method is equivalent to the union of
the asymmetric parts of $T\left( R_{\alpha }\right) $ for all $\alpha \in %
\left[ 1/2,1\right) $.

By Proposition 4 and Corollary 2, we immediately have the following result.

\begin{corollary}
\begin{equation*}
\mathcal{T}=\mathcal{R}\left( \bigcup_{\alpha \in \left[ 1/2,1\right)
}P\left( T\left( R_{\alpha }\right) \right) \right) =\mathcal{R}\left(
R^{Sc}\right) \text{.}
\end{equation*}
\end{corollary}

This result can be rewritten as $R\in \mathcal{R}\left( R^{Sc}\right) $ if
and only if $R\in \mathcal{R}\left( T\left( R_{\alpha }\right) \right) $ for
all $\alpha \in \left[ 1/2,1\right) $. Thus, any linear order that respects
the binary relation constructed by the Schulze method also respects the
transitive closure of the binary relation constructed by any supermajority
rules. Moreover, any linear order that respects the transitive closures of
all supermajority binary relations also respects the binary relation
constructed by the Schulze method.

By Propositions 4 and 5, and Corollary 2, the linear orders that are derived
by the Schulze method and the ranked pairs method are in $\mathcal{S}$.
Thus, both methods belong to the class of those that select one from $%
\mathcal{S}$. However, any linear orders that are derived by the ranked
pairs method may not be in $\mathcal{T}$. Furthermore, there may be some
linear orders in $\mathcal{S}$ that are derived by neither of them. We
illustrate these observations with an example in Appendix C.

Since $R^{Sc}$ is a partial order, we have the following result.

\begin{corollary}
$\mathcal{T}\neq \emptyset .$
\end{corollary}

By Proposition 5 and Corollary 3, the $T$-order set is a refinement of the $%
S $-order set.

\section{Properties}

In this section, we consider the properties of collective linear orders.
First, we consider the property introduced by Truchon (1998) and called the
extended Condorcet criterion.

Let $\mathcal{A}$ be the set of all partitions of $A$ such that for all 
\begin{equation*}
\mathbf{A}=\left\{ A_{1},\cdots ,A_{X}\right\} \in \mathcal{A},
\end{equation*}
all $A_{x},A_{x^{\prime }}\in \mathbf{A}$ with $x<x^{\prime }$, all $a\in
A_{x}$ and all $a^{\prime }\in A_{x^{\prime }}$, $N\left[ a,a^{\prime }%
\right] >N\left[ a^{\prime },a\right] $. A linear order $R^{\prime }$ is
said to satisfy the\textbf{\ extended Condorcet criterion} if for all $%
\mathbf{A}\in \mathcal{A},$ all $A_{x},A_{x^{\prime }}\in \mathbf{A}$ with $%
x<x^{\prime }$, all $a\in A_{x}$ and all $a^{\prime }\in A_{x^{\prime }}$, $%
\left( a,a^{\prime }\right) \in R^{\prime }$.

We introduce the following terms with regards to $R_{1/2}$. An alternative $%
a $ is said to be the \textbf{Condorcet winner} if $\left( a,b\right) \in
R_{1/2}$ for all $b\in A\setminus \left\{ a\right\} $. On the other hand, $a$
is said to be the \textbf{Condorcet loser} if $\left( b,a\right) \in R_{1/2}$
for all $b\in A\setminus \left\{ a\right\} $. Suppose that $R^{\prime
}:a_{1},\cdots ,a_{\left\vert I\right\vert }$ is a linear order that
satisfies the\textbf{\ }extended Condorcet criterion. Then, $a_{1}$ is the
Condorcet winner if there is one, and $a_{\left\vert I\right\vert }$ is the
Condorcet loser if there is one.

Truchon (1998) shows that any linear order that is derived by the
Kemeny-Young method also satisfies the\textbf{\ }extended Condorcet
criterion for any $\mathbf{R\in }\mathcal{W}^{\left\vert I\right\vert }$.

We have the following result.

\begin{proposition}
If $R\in \mathcal{R}\left( S\left( R_{1/2}\right) \right) ,$ then $R$
satisfies the\textbf{\ }extended Condorcet criterion.
\end{proposition}

We immediately have the following result.

\begin{corollary}
Any $R\in \mathcal{S}$ satisfies the\textbf{\ }extended Condorcet criterion.
\end{corollary}

Roughly speaking, the extended Condorcet criterion requires that the linear
order must respect $R_{1/2}$. Since a linear order in $\mathcal{S}$ must
respect $S\left( R_{1/2}\right) $, we have this result.

Note that there is $R\notin \mathcal{R}\left( S\left( R_{1/2}\right) \right) 
$ that satisfies the\textbf{\ }extended Condorcet criterion. To show this
fact, we revisit Example 2. In this case, any linear order satisfies the%
\textbf{\ }extended Condorcet criterion. However, some linear orders such as 
$R:$ $d,a,b,c$, are not included in $\mathcal{R}\left( S\left(
R_{1/2}\right) \right) $, because $\left( a,d\right) ,\left( b,d\right) \in
P\left( S\left( R_{1/2}\right) \right) $.\footnote{%
Ando et al. (2022) introduce a stronger property than the extended Condorcet
criterion called the strong Condorcet criterion. However, Proposition 7
continues to hold even if the\textbf{\ }extended Condorcet criterion is
replaced by the strong Condorcet criterion. Moreover, there also is $R\notin 
\mathcal{R}\left( S\left( R_{1/2}\right) \right) $ that satisfies the\textbf{%
\ }strong Condorcet criterion. In fact, in Example 2, $R:$ $d,a,b,c$
satisfies strong Condorcet criterion, but is not included in $\mathcal{R}%
\left( S\left( R_{1/2}\right) \right) $.}

Next, we consider the Pareto principle. A linear order $R^{\prime }$ is said
to satisfy the\textbf{\ Pareto principle} if for any $a,b\in A$ such that $%
\left( b,a\right) \notin P\left( R^{i}\right) $ for all $i\in I$ and $\left(
a,b\right) \in P\left( R^{i}\right) $ for some $i\in I$, $\left( a,b\right)
\in R^{\prime }$.\footnote{%
There are several kinds of the Pareto principle. Our definition is in the
spirit of (i) of the \textquotedblleft strong Pareto\textquotedblright\
defined by Bossert and Suzumura (2010, P.129). Schulze (2018) also considers
the Pareto principle, but his definitions are different from ours.}

We consider the two cases. First, suppose that each individual has a linear
order preference.

\begin{proposition}
Suppose that $R^{i}$ is a linear order for all $i\in I$. For any $\hat{R}\in 
\mathcal{R}\left( S\left( R_{\alpha }\right) \right) $ for sufficiently
large $\alpha \in \left[ 1/2,1\right) $, $\hat{R}$ satisfies the\textbf{\ }%
Pareto principle.
\end{proposition}

Second, suppose that each individual has a weak order preference. In this
case, some additional assumption is necessary.

\begin{proposition}
Suppose that $R^{i}$ is a weak order for all $i\in I$, and $f\left(
n,0\right) =1>f\left( m,1\right) $ for any $n>0$ and any $m\geq 0$. For any $%
\hat{R}\in \mathcal{R}\left( S\left( R_{\alpha }\right) \right) $ for
sufficiently large $\alpha \in \left[ 1/2,1\right) $, $\hat{R}$ satisfies the%
\textbf{\ }Pareto principle.
\end{proposition}

Thus, we have the following result.

\begin{corollary}
Suppose that $R^{i}$ is a linear order for all $i\in I$, or $R^{i}$ is a
weak order for all $i\in I$ and $f\left( n,0\right) =1>f\left( m,1\right) $
for any $n>0$ and any $m\geq 0$. Then, $\hat{R}\in \mathcal{S}$ satisfies the%
\textbf{\ }Pareto principle.
\end{corollary}

Note that $f\left( n,0\right) =1>f\left( m,1\right) $ is satisfied if (\ref%
{b}) is used, but it is not satisfied if (\ref{b1}) is used. We provide an
example with (\ref{b1}) (without $f\left( n,0\right) =1>f\left( m,1\right) $%
) such that some linear orders on $A$ in both $\mathcal{T}$ and $\mathcal{S}$
do not satisfy the Pareto principle.

\subsubsection*{Example 3}

There are $A=\left\{ a,b,c,d\right\} $ and $I=\left\{ 1,2,3\right\} $. All
individuals have a weak order preference on $A$ and%
\begin{equation*}
R^{1}:\text{ }\left[ a,d\right] ,b,c,\text{ }R^{2}:\text{ }b,c,\left[ a,d%
\right] ,\text{ }R^{3}:\text{ }c,a,d,b,
\end{equation*}%
meaning that $a$ and $d$ are indifferent alternatives for individuals $1$
and $2$. Note that $\left( d,a\right) \notin P\left( R^{i}\right) $ for all $%
i\in I$ and $\left( a,d\right) \in P\left( R^{3}\right) $. Thus, for any
linear order $\hat{R}$ that satisfies the Pareto principle, $\left(
a,d\right) \in \hat{R}$.

Suppose that (\ref{b1}) holds. Then, for $\alpha \in \left[ 1/2,2/3\right) $
and $\alpha ^{\prime }\in \left[ 2/3,1\right) ,$ 
\begin{eqnarray*}
R_{\alpha } &=&\left\{ \left( a,d\right) ,\left( a,b\right) ,\left(
d,b\right) ,\left( c,a\right) ,\left( c,d\right) ,\left( b,c\right) \right\}
, \\
P\left( T\left( R_{\alpha }\right) \right)  &=&P\left( S\left( R_{\alpha
}\right) \right) =\emptyset , \\
R_{\alpha ^{\prime }} &=&T\left( R_{\alpha ^{\prime }}\right) =S\left(
R_{\alpha ^{\prime }}\right) =\emptyset ,
\end{eqnarray*}%
Any linear orders on $A$ are in both $\mathcal{T}$ and $\mathcal{S}$. Thus,
some linear orders on $A$ in both $\mathcal{T}$ and $\mathcal{S}$ such as $%
\hat{R}:$ $d,c,b,a$ do not satisfy the Pareto principle.

On the other hand, suppose that (\ref{b}) holds. Then, for $\alpha \in \left[
1/2,2/3\right) $ and $\alpha ^{\prime }\in \left[ 2/3,1\right) ,$ 
\begin{eqnarray*}
R_{\alpha } &=&\left\{ \left( a,d\right) ,\left( a,b\right) ,\left(
d,b\right) ,\left( c,a\right) ,\left( c,d\right) ,\left( b,c\right) \right\}
, \\
P\left( T\left( R_{\alpha }\right) \right)  &=&P\left( S\left( R_{\alpha
}\right) \right) =\emptyset , \\
R_{\alpha ^{\prime }} &=&T\left( R_{\alpha ^{\prime }}\right) =S\left(
R_{\alpha ^{\prime }}\right) =\left\{ \left( a,d\right) \right\} ,
\end{eqnarray*}%
Thus, in this case, any linear order $\hat{R}\in \mathcal{S}\left( =\mathcal{%
T}\right) $ satisfies $\left( a,a^{\prime }\right) \in \hat{R}$.

\section{Concluding Remarks}

Our approach for obtaining a linear order is related to two influential
methods: Tideman's ranked pairs method and the Schulze method. These methods
are examples of Condorcet methods, which achieve linear orders possessing
the same characteristics described in Corollary 4. Although Condorcet
methods have received considerable attention in voting theory research, they
are not widely applied in real-world scenarios. A primary reason for this
may be the difficulty in making general voters understand methods such as
ranked pairs and the Schulze method. \footnote{%
For example, Poundstone (2008, Ch.13) mentions that \textquotedblleft Maybe
the biggest hitch with Condorcet voting is the simplicity issue. Condorcet
supporters talk up the populist appeal of the John Wayne, last-man-standing
premise. Soccer moms and NASCAR dads will nod their heads to that. The thing
is, its impossible to explain Condorcet voting any further without talking
about cycles and \textquotedblleft beat-paths\textquotedblright\ and
\textquotedblleft cloneproof Schwartz sequential
dropping.\textquotedblright\ This is where the androids mask falls off and
its all wires and microchips inside. Non-Ph.D.s run screaming for the exits.
So far, Condorcet voting has tended to appeal to the kind of people who can
write Javascript code for it.\textquotedblright} Our results provide
intuitive interpretations of these two methods, potentially making them
easier to explain. The interpretation via supermajority rules may especially
help enhance comprehensibility.

Finally, we address several remaining issues. First, we provide some remarks
on applying our approach when selecting only one winner. The result
indicates that establishing the non-emptiness of the $S$-order set
(Corollary 1) is delicate. Second, we introduce a computationally efficient
method for deriving the $S$-order set.

\subsection{Single winner election}

In this subsection, we focus on methods to have only one winner in election.
First, we consider the \textbf{Simpson-Kramer min-max method}, which is in a
similar spirit to that considered in 4.1. We try to refine the set of
winners of this method, which is calculated as 
\begin{equation*}
W^{SK}=\arg \min_{a\in A}\left[ \max_{b\in A}\left[ \Phi \left[ b,a\right] %
\right] \right] \text{.}
\end{equation*}%
Then, we have the following result.

\begin{remark}
\begin{equation}
W^{SK}=\bigcap_{\alpha \in \left[ 1/2,1\right) :\text{ }M(A,R_{\alpha })\neq
\emptyset }M(A,R_{\alpha })\text{.}  \label{g}
\end{equation}
\end{remark}

In words, the intersection of nonempty maximal sets for $\alpha \in \left[
1/2,1\right) $ is equivalent to the set of winners of the Simpson-Kramer
min-max method. However, any element of the intersection may be a Condorcet
loser. To show this, we reexamine Example 1. In this example%
\begin{eqnarray}
M(A,R_{\alpha }) &=&\emptyset \text{ if }\alpha \in \left[ 1/2,5/9\right) 
\notag \\
&=&\left\{ d\right\} \text{ if }\alpha \in \left[ 5/9,2/3\right)  \label{x}
\\
&=&\left\{ a,b,c,d\right\} \text{ if }\alpha \in \left[ 2/3,1\right) \text{.}
\notag
\end{eqnarray}%
Therefore, 
\begin{equation*}
W^{SK}=\bigcap_{\alpha \in \left[ 1/2,1\right) :\text{ }M(A,R_{\alpha })\neq
\emptyset }M(A,R_{\alpha })=\left\{ d\right\} \text{.}
\end{equation*}%
As is mentioned above, $d$ is the Condorcet loser but the unique winner of
the Simpson-Kramer min-max method. Since $d$ is the the unique winner,
contrary to the analysis in Section 4.1, we cannot refine $W^{SK}$ by using
a similar method to that introduced in this study. That is, we have the
following result.

\begin{remark}
\begin{equation}
\left( \bigcap\limits_{\alpha \in \left[ 1/2,1\right) :\text{ }M(A,R_{\alpha
})\neq \emptyset }M(A,R_{\alpha })\right) \bigcap \left(
\bigcap\limits_{\alpha \in \left[ 1/2,1\right) :\text{ }M(A,R_{\alpha
})=\emptyset }M(A,S\left( R_{\alpha }\right) )\right)  \label{z}
\end{equation}%
may be empty.
\end{remark}

We use $R_{\alpha }$ for all $\alpha \in \left[ 1/2,1\right) $ but modify it
when the maximal sets for $R_{\alpha }$ is empty. In some cases, we cannot
choose any winner. We reexamine Example 1 to show this fact. If $\alpha \in %
\left[ 1/2,5/9\right) $, $M(A,R_{\alpha })=\emptyset $ and $M(A,S\left(
R_{\alpha }\right) )=\left\{ a,b,c,e\right\} $. By (\ref{x}), in this
example, (\ref{z}) is empty. This implies that we cannot refine the
Simpson-Kramer min-max winner by this approach.

Next, we reconsider the Schulze method and the ranked pairs method, which
are introduced as methods determining the single winner. The set of winners
of Schulze method is $W^{Sc}=M(A,R^{Sc})$ and that of the ranked pairs
method is 
\begin{equation*}
W^{RP}=\left\{ a\in A\text{ }\left\vert \text{ }R^{\pi }\left( 1\right) =a%
\text{ for some }\pi \in \Pi \right. \right\} .
\end{equation*}

We have the following results.

\begin{corollary}
\begin{equation*}
W^{Sc}=\bigcap\limits_{\alpha \in \left[ 1/2,1\right) }M\left( A,T\left(
R_{\alpha }\right) \right) =M\left( A,\bigcup_{\alpha \in \left[
1/2,1\right) }P\left( T\left( R_{\alpha }\right) \right) \right) .
\end{equation*}
\end{corollary}

This implies that the set of winners of the Schulze method is equivalent to
the intersection of the maximal sets for $T\left( R_{\alpha }\right) $ for
all $\alpha \in \left[ 1/2,1\right) $. Since $R_{\alpha }$ is asymmetric for
all $\alpha \in \left[ 1/2,1\right) $ (the first result of Lemma 8), $%
M\left( A,T\left( R_{\alpha }\right) \right) $ is equivalent to the Schwartz
set for $R_{\alpha }$. Therefore, $W^{Sc}$ is the intersection of the
Schwartz sets for $R_{\alpha }$ for all $\alpha \in \left[ 1/2,1\right) $.

\begin{corollary}
\begin{equation}
W^{Sc}\cup W^{RP}\subseteq \bigcap\limits_{\alpha \in \left[ 1/2,1\right)
}M\left( A,S\left( R_{\alpha }\right) \right) =M\left( A,\bigcup_{\alpha \in %
\left[ 1/2,1\right) }P\left( S\left( R_{\alpha }\right) \right) \right) .
\label{y1}
\end{equation}
\end{corollary}

This implies that the sets of winners of the Schulze method and the ranked
pairs method are subsets of the intersection of the maximal sets for $%
S\left( R_{\alpha }\right) $ for all $\alpha \in \left[ 1/2,1\right) $.

Note that (\ref{z}) and (\ref{y1}) may be different even though $S\left(
R_{\alpha }\right) =R_{\alpha }$ if $R_{\alpha }$ is P-acyclic. This
discrepancy occurs because $M(A,R_{\alpha })\neq \emptyset $ may be
satisfied even if $R_{\alpha }$ has some P-cyclic. For example, if $\alpha
\in \left[ 5/9,2/3\right) $, then $R_{\alpha }$ has some cycles (such as $%
a\rightarrow b\rightarrow c\rightarrow a$), but $d\in M(A,R_{\alpha })$. In
this case, $S\left( R_{\alpha }\right) \neq R_{\alpha }$. Therefore, (\ref{z}%
) may be empty, but (\ref{y1}) must be nonempty. As illustrated by this
conclusion, the establishment of the non-emptiness of $\mathcal{S}$
(Corollary 1) is rather delicate.

\subsection{Method to Derive Linear Orders}

Finally, we provide methods to derive the $S$-order set and the $T$-order
set, respectively.

First, by Proposition 6, the construction method of $\bigcup_{\alpha \in %
\left[ 1/2,1\right) }P\left( T\left( R_{\alpha }\right) \right) $ is the
same as the Schulze rule. Therefore, we can use the method introduced by
Schulze (2011). Here, we provide a method to construct $\bigcup_{\alpha \in %
\left[ 1/2,1\right) }P\left( S\left( R_{\alpha }\right) \right) $.

\begin{proposition}
$\left( a,b\right) \in \bigcup_{\alpha \in \left[ 1/2,1\right) }P\left(
S\left( R_{\alpha }\right) \right) $ if and only if 
\begin{equation*}
\Phi \left[ a,b\right] >\max \left\{ \frac{1}{2},B[b,a]\right\} .
\end{equation*}
\end{proposition}

Since we can have $B[a,b]$ for all $a,b\in A$ by using Dijkstra's algorithm,
we can computationally efficiently construct $\bigcup_{\alpha \in \left[
1/2,1\right) }P\left( S\left( R_{\alpha }\right) \right) $.

By using $\bigcup_{\alpha \in \left[ 1/2,1\right) }P\left( S\left( R_{\alpha
}\right) \right) $ and $\bigcup_{\alpha \in \left[ 1/2,1\right) }P\left(
T\left( R_{\alpha }\right) \right) $, we can derive the $S$-order set and
the $T$-order set by using the class of algorithms in the previous section,
because of Theorems 1 and 2.

\section*{References}

\begin{description}
\item Ando, K., Sukegawa, N., Takagi, S. (2022) Strong Condorcet Criterion
for the linear ordering, Journal of the Operations Research Society of Japan
65(2), 67--75

\item Arrow, K.J. (1963) Social Choice and Individual Values. Yale
University Press, New Haven, CT.

\item Arrow, K.J. (1977) Current Developments in the Theory of Social
Choice,\ Social Research 44, 607-622.

\item Balasko, Y., Cr\`{e}s H., (1997) The probability of Condorcet cycles
and super majority rules,\ Journal of Economic Theory 75, 237-270.

\item Barber\'{a}, S., Bossert W. (2023) Opinion aggregation: Borda and
Condorcet revisited, Journal of Economic Theory 210, 105654.

\item Black, D. (1958) The Theory of Committees and Elections. Cambridge
University Press, London.

\item Borda, J.-C. (1781) Memoires sue les Elections au Scrutin,\ Histoires
de I'Academie Royale des Sciences. English translation by A. de Grazia.
Mathematical Derivation of an Election System,\ Isis 44 (1953).

\item Bordes, G. (1976) Consistency, Rationality, and Collective Choice,\
Review of Economic Studies 43, 451-457.

\item Bossert, W., Sprumont Y., Suzumura, K. (2005) Consistent
Rationalizability,\ Economica 72, 85--200.

\item Bossert, W., Suzumura, K. (2010) Consistency, Choice, and Rationality,
Harvard University Press.

\item Caplin, A., Nalebuff, B. (1988) On 64\% majority rule,\ Econometrica
56, 784-814.

\item Coughlin, P.J. (1981) Necessary and Sufficient Conditions for $\delta $%
-relative Majority Voting Equilibria.\ Econometrica 49(5), 1223--1224

\item Condorcet, M. (1785) Essai Sur L'Application de L'Analyse a la
Probabilite Des Decisions Rendues a la Pluralite Des Voix.\ Paris.

\item Cr\`{e}s H., \"{U}nver, M. U. (2017) Toward a 50\%-majority
Equilibrium When Voters are Symmetrically Distributed, Mathematical Social
Sciences 90, 145-149.

\item Deb, R. (1977) On Schwartz's Rule.\ Journal of Economic Theory 16,
103-110.

\item Fujiwara-Greve, T., Kawada, Y., Nakamura, Y., Okamoto, N., 2023.
Accountable Voting. Available at SSRN: https://ssrn.com/abstract=4406802

\item Gjorgjiev, R., Xefteris, D. (2015) Transitive Supermajority Rule
Relations,\ Economics Theory Bulletin 3, 299--312.

\item Green-Armytage, J., Tideman, N. (2020) Selecting the Runoff-pair,\
Public Choice 182, 119--137

\item Greenberg, J. (1979) Consistent majority rule over compact sets of
alternatives,\ Econometrica 47, 627-663.

\item Nakamura, K. (1979) The voters in a simple game with ordinal
preferences,\ International Journal of Game Theory 8, 55--61.

\item Pattanaik, P.K. (2002) Positional Rules of Collective
Decision-making,\ in Handbook of Social Choice and Welfare. Vol. 1, Edited
by Arrow, K.J. Sen A.K., and Suzumura K., North-Holland. 361-394.

\item Ohbo, K., Tsurutani, M., Umezawa, M., Yamamoto, Y. (2005) Social
welfare function for restricted individual preference domain, Pacific
Journal of Optimization 1(2), 315-325.

\item Okumura, Y. (2024) Aggregating Incomplete Rankings, Available at
https://arxiv.org/html/2402.16309v3

\item Saari, D.G. (1994) Geometry of Voting, Springer Nature.

\item Schwartz, T. (1972) Rationality and the Myth of the Maximum.\ Nous 6,
97-117.

\item Schulze, M. (2011) A New Monotonic, Clone-independent, Reversal
Symmetric, and Condorcet-consistent Single-winner Election Method, Social
Choice and Welfare 36(2), 267--303.

\item Slutsky, S. (1979) Equilibrium under $\alpha $-majority voting,\
Econometrica 47(5), 1113--1125.

\item Suzumura, K. (1983) Rational Choice, Collective Decisions and Social
Welfare, New York: Cambridge University Press.

\item Suzumura, K. (2012) Introduction to Social Choice Theory (in
Japanese), Tokyo: Toyo Keizai Inc.

\item Tideman, T. N. (1987) Independence of Clones as a Criterion for Voting
Rules, Social Choice and Welfare 4(3), 185--206.

\item Tideman, T. N. (2006) Collective Decisions and Voting: The Potential
for Public Choice, Ashgate Publishing.

\item Tovey, C. A. (1997) Probabilities of Preferences and Cycles with Super
Majority Rules, Journal of Economic Theory 75, 271-279.

\item Truchon, M. (1998) An Extension of the Concordet Criterion and Kemeny
Orders, Technical Report, cahier 98-15 du Centre de Recherche en \'{E}%
conomie et Finance Appliqu\'{e}es, Universit\'{e} Laval, Qu\'{e}bec, Canada.

\item Young, H. P. (1988) Condorcet's theory of voting, American Political
Science Review 82(4), 1231--1244.

\item Villar, A., (2023) The precedence function: a numerical evaluation
method for multi criteria ranking problems, Economic Theory Bulletin 11,
211--219.
\end{description}

\section*{Appendix A: Technical Results}

To show our results, we use the following technical results.

\begin{lemma}
\begin{enumerate}
\item If $R$ is transitive, then it is Suzumura consistent.

\item If $R$ is Suzumura consistent, then it is P-acyclic.

\item If $R$ is asymmetric, then it is P-acyclic if and only if it is
Suzumura consistent.
\end{enumerate}
\end{lemma}

Lemma 1 follows from Suzumura (1983) and Bossert and Suzumura (2010).

\begin{lemma}
If $R$ is P-acyclic, then $M\left( A^{\prime },\succ _{s}\right) $ is
non-empty for any finite $A^{\prime }\subseteq A$.
\end{lemma}

Lemma 2 follows from Bossert and Suzumura (2010, Theorem 2.4).

\begin{lemma}
Let $R_{1},\cdots ,R_{J}$ be binary relations. Then, 
\begin{equation}
\bigcap\limits_{j\in \left\{ 1,\cdots ,J\right\} }\mathcal{R}\left(
R_{j}\right) \subseteq \mathcal{R}\left( \bigcup\limits_{j\in \left\{
1,\cdots ,J\right\} }P\left( R_{j}\right) \right) .  \label{a}
\end{equation}
\end{lemma}

\textbf{Proof.} If the left-hand side of (\ref{a}) is empty, then this is
trivial. Thus, we assume that it is not empty and let $\hat{R}$ be an
element in it. Suppose not; that is, $\hat{R}$ not an element in the
right-hand of (\ref{a}). Then, there are $a,b\in A$ such that $\left(
a,b\right) \in \hat{R}$ but $\left( b,a\right) \in \bigcup\nolimits_{j\in
\left\{ 1,\cdots ,J\right\} }P\left( R_{i}\right) $. This implies that $%
\left( b,a\right) \in P\left( R_{j}\right) ;$ that is, $\left( b,a\right)
\in R_{j}$ and $\left( a,b\right) \notin R_{j}$ for some $j=1,\cdots ,J$.
However, these imply $\hat{R}\notin \mathcal{R}\left( R_{j}\right) $, which
is a contradiction. \textbf{Q.E.D.}

\begin{lemma}
If $R_{1}\subseteq R_{2}\subseteq \cdots \subseteq R_{J}$ and $R_{j}$ is
P-acyclic for all $j\in \left\{ 1,\cdots ,J\right\} $, then 
\begin{equation*}
\bigcap\limits_{j\in \left\{ 1,\cdots ,J\right\} }\mathcal{R}\left(
R_{j}\right) =\mathcal{R}\left( \bigcup\limits_{j\in \left\{ 1,\cdots
,J\right\} }P\left( R_{j}\right) \right) .
\end{equation*}
\end{lemma}

\textbf{Proof. }By Lemma 3, it is sufficient to show 
\begin{equation}
\bigcap\limits_{j\in \left\{ 1,\cdots ,J\right\} }\mathcal{R}\left(
R_{j}\right) \supseteq \mathcal{R}\left( \bigcup\limits_{j\in \left\{
1,\cdots ,J\right\} }P\left( R_{j}\right) \right) .  \label{c}
\end{equation}%
If the right-hand side of (\ref{c}) is empty, (\ref{c}) is trivial. Thus,
suppose that $\hat{R}$ is an element of the right-hand side of (\ref{c}).
Then, toward a contradiction, suppose that $\hat{R}$ is not an element of
the left-hand side of (\ref{c}). By the P-acyclicity of $R_{j}$ for all $%
j\in \left\{ 1,\cdots ,J\right\} $ and Proposition 1, $\mathcal{R}\left(
R_{j}\right) \neq \emptyset $ for all $j\in \left\{ 1,\cdots ,J\right\} $.
Therefore, there are $a,b\in A$ and $j^{\prime }\in \left\{ 1,\cdots
,J\right\} $ such that $\left( a,b\right) \in \hat{R}$, $\left( b,a\right)
\in R_{j^{\prime }}$ and $\left( a,b\right) \notin R_{j^{\prime }}$. Since $%
\hat{R}$ is asymmetric, $\left( b,a\right) \notin \hat{R}$. Since $\left(
b,a\right) \in P\left( R_{j^{\prime }}\right) $ and $\left( a,b\right) \in 
\hat{R}$, for some $j^{\prime \prime }\in \left\{ 1,\cdots ,J\right\} $, $%
\left( a,b\right) \in P\left( R_{j^{\prime \prime }}\right) $; that is, $%
\left( a,b\right) \in R_{j^{\prime \prime }}$ and $\left( b,a\right) \notin
R_{j^{\prime \prime }}$. If $R_{j^{\prime \prime }}\subseteq R_{j^{\prime }}$%
, then $\left( a,b\right) \in R_{j^{\prime \prime }}$ implies $\left(
a,b\right) \in R_{j^{\prime }}$, which is a contradiction. Thus, $%
R_{j^{\prime }}\subseteq R_{j^{\prime \prime }}$, but $\left( b,a\right) \in
R_{j^{\prime }}$ implies $\left( b,a\right) \in R_{j^{\prime \prime }}$,
which is a contradiction. Therefore, we have Lemma 4. \textbf{Q.E.D.}

\begin{lemma}
$P\left( S\left( R\right) \right) \subseteq P\left( T\left( R\right) \right) 
$ and $P\left( S\left( R\right) \right) \subseteq P\left( R\right) $.
\end{lemma}

\textbf{Proof. }Let $\left( a,b\right) \in P\left( S\left( R\right) \right) $%
; that is, $\left( a,b\right) \in S\left( R\right) $ and $\left( b,a\right)
\notin S\left( R\right) $. Since $S\left( R\right) \subseteq T\left(
R\right) $, $\left( a,b\right) \in T\left( R\right) $.

First, we show $P\left( S\left( R\right) \right) \subseteq P\left( T\left(
R\right) \right) $. Suppose not; that is, $\left( a,b\right) \notin P\left(
T\left( R\right) \right) $. Since $\left( a,b\right) \in T\left( R\right) $, 
$\left( b,a\right) \in T\left( R\right) $. Then, $\left( b,a\right) \in
T\left( R\right) $ and $\left( b,a\right) \notin S\left( R\right) $ imply $%
\left( a,b\right) \notin R$. Then, $\left( a,b\right) \in T\left( R\right) $
and $R\subseteq T\left( R\right) $ are not compatible. Hence $P\left(
S\left( R\right) \right) \subseteq P\left( T\left( R\right) \right) $.

Second, we show $P\left( S\left( R\right) \right) \subseteq P\left( R\right) 
$. Since $\left( b,a\right) \notin S\left( R\right) $, $\left( b,a\right)
\notin R$. Moreover, since $\left( b,a\right) \notin R$ and $\left(
a,b\right) \in S\left( R\right) $, $\left( a,b\right) \in R$. Therefore, $%
\left( a,b\right) \in P\left( R\right) $. \textbf{Q.E.D.}\newline

\begin{lemma}
If $R\subseteq R^{\prime }$, then $T\left( R\right) \subseteq T\left(
R^{\prime }\right) $ and $S\left( R\right) \subseteq S\left( R^{\prime
}\right) $.
\end{lemma}

For the proof of Lemma 6, see Suzumura (1983) and Bossert and Suzumura
(2010).

\begin{lemma}
\begin{enumerate}
\item For all $\alpha \in \left[ 1/2,1\right) $, $R_{\alpha }$ is asymmetric.

\item For all $\alpha ,\beta \in \left[ 1/2,1\right) $ such that $\alpha
\geq \beta $, $R_{\alpha }\subseteq R_{\beta }$.

\item For all $\alpha ,\beta \in \left[ 1/2,1\right) $ such that $\alpha
\geq \beta $, if $R_{\beta }$ is P-acyclic, then $R_{\alpha }$ is also
P-acyclic, and moreover, if $R_{\alpha }$ is not P-acyclic, then $R_{\beta }$
is also not P-acyclic.

\item $\left( a,b\right) \in R_{1/2}$ if and only if $N\left[ a,b\right] >N%
\left[ b,a\right] $
\end{enumerate}
\end{lemma}

\textbf{Proof.} Fix $\alpha ,\beta $ such that $1/2\leq \beta \leq \alpha <1$%
. We show the first result. Suppose $\left( a,b\right) \in R_{\alpha }$.
Then, $\Phi \left[ a,b\right] =f\left( N\left[ a,b\right] ,N\left[ b,a\right]
\right) >\alpha \geq 1/2$. Since $f\left( N\left[ b,a\right] ,N\left[ b,a%
\right] \right) =1/2$ and $f\left( n,N\left[ b,a\right] \right) \leq f\left(
N\left[ b,a\right] ,N\left[ b,a\right] \right) $ for all $n\leq N\left[ b,a%
\right] $, we have $N\left[ a,b\right] >N\left[ b,a\right] $. Then, $\Phi %
\left[ b,a\right] =f\left( N\left[ b,a\right] ,N\left[ a,b\right] \right)
\leq f\left( N\left[ b,a\right] ,N\left[ b,a\right] \right) $ and $f\left( N%
\left[ b,a\right] ,N\left[ a,b\right] \right) \leq 1/2\leq \alpha $. Hence, $%
\left( a,b\right) \notin R_{\alpha }$. Therefore, $R_{\alpha }$ is
asymmetric.

We show the second result. If $\left( a,b\right) \in R_{\alpha },$ then $%
\Phi \left[ a,b\right] >\alpha \geq \beta $ and thus $\left( a,b\right) \in
R_{\beta }$.

We show the third result. First, a subset of a P-acyclic binary relation is
also P-acyclic. Therefore, by the third result, we have the latter part of
the third result. Second, suppose $R_{\alpha }$ has a P-cycle. By the first
and second results, $R_{\beta }$ is asymmetric and $R_{\alpha }\subseteq
R_{\beta }$. Therefore, $R_{\beta }$ also has that P-cycle and, thus we have
the former part of the third result.

We show the fourth result. First, suppose $N\left[ a,b\right] >N\left[ b,a%
\right] $. Then, 
\begin{equation*}
\Phi \left[ a,b\right] =f\left( N\left[ a,b\right] ,N\left[ b,a\right]
\right) >f\left( N\left[ b,a\right] ,N\left[ b,a\right] \right) =1/2\text{.}
\end{equation*}%
Second, suppose $N\left[ a,b\right] \leq N\left[ b,a\right] $. Then, 
\begin{equation*}
\Phi \left[ a,b\right] =f\left( N\left[ a,b\right] ,N\left[ b,a\right]
\right) \leq f\left( N\left[ b,a\right] ,N\left[ b,a\right] \right) =1/2%
\text{.}
\end{equation*}%
Therefore, we have the fourth result. \textbf{Q.E.D.}

\begin{lemma}
Suppose $B[b,a]\in \left[ 1/2,1\right) $. First, $\mathcal{B}\left[ \left(
b,a\right) ,R_{B[b,a]}\right] =\emptyset $. Second, if $B[a,b]>B[b,a]$, then 
$\mathcal{B}\left[ \left( a,b\right) ,R_{B[b,a]}\right] \neq \emptyset $.
\end{lemma}

\textbf{Proof. }Suppose $B[b,a]\in \left[ 1/2,1\right) $. First, let $\left(
a_{1},a_{2},\cdots ,a_{m}\right) $ be an arbitrary sequence of distinct
alternatives such that $a_{1}=b$ and $a_{m}=a$. By the definition of $B[b,a]$%
, $B[b,a]\geq \Phi \left[ a_{j},a_{j+1}\right] $ for some $j=1,\cdots ,m-1$.
Therefore, $\left( a_{1},a_{2},\cdots ,a_{m}\right) \notin \mathcal{B}\left[
\left( b,a\right) ,R_{B[b,a]}\right] $ and thus $\mathcal{B}\left[ \left(
b,a\right) ,R_{B[b,a]}\right] =\emptyset $.

Second, suppose $B[a,b]>B[b,a]$. By the definition of $B[a,b]$, there is a
sequence of distinct alternatives $\left( a_{1},a_{2},\cdots ,a_{m}\right) $
such that $a_{1}=a$, $a_{m}=b$ and $\Phi \left[ a_{j},a_{j+1}\right] \geq
B[a,b]>B[b,a]$ for all $j=1,\cdots ,m-1$. Then, $\left( a_{1},a_{2},\cdots
,a_{m}\right) \in \mathcal{B}\left[ \left( b,a\right) ,R_{B[b,a]}\right] $.
Thus, $\mathcal{B}\left[ \left( a,b\right) ,R_{B[b,a]}\right] \neq \emptyset 
$. \textbf{Q.E.D.}

\begin{lemma}
If $R^{i}$ is weak order for all $i\in I$ and $\alpha \in \left[
1/2,1\right) $ is sufficiently large, then $R_{\alpha }$ is P-acyclic.
\end{lemma}

\textbf{Proof.} We fix $\alpha \in \left[ f\left( \left\vert I\right\vert
-1,1\right) ,1\right) $. Then, $\left( a,b\right) \in R_{\alpha }$ implies $%
f\left( N\left[ a,b\right] ,N\left[ b,a\right] \right) >f\left( \left\vert
I\right\vert -1,1\right) $. Therefore, $N\left[ a,b\right] =\left\vert
I\right\vert $ and/or $N\left[ b,a\right] =0$. If $N\left[ a,b\right]
=\left\vert I\right\vert $; that is, if all individuals prefer $a$ to $b$,
then $N\left[ b,a\right] =0$. Therefore, $\left( a,b\right) \in R_{\alpha }$
implies $N\left[ b,a\right] =0$; that is, no individual prefers $b$ to $a$.
Moreover, since $f\left( 0,0\right) =1/2$, $N\left[ a,b\right] \geq 1$; that
is, at least one individual prefers $a$ to $b$.

Now, we show that $R_{\alpha }$ is P-acyclic. Suppose that there is $\left(
a_{0},a_{1},\cdots ,a_{m}\right) \in \mathcal{B}\left[ \left( k,j\right)
,P\left( R_{\alpha }\right) \right] $. We show $\left( a_{K},a_{0}\right)
\notin P\left( R_{\alpha }\right) $. Since $\left( a_{0},a_{1}\right) \in
R_{\alpha }$, there is at least one individual $i$ such that $\left(
a_{0},a_{1}\right) \in R^{i}$. Moreover, since $N\left[ b,a\right] =0$, $%
(a_{l},a_{l-1})\notin P\left( R^{i}\right) $ for all $l=1,\cdots ,m$. We
show $\left( a_{0},a_{2}\right) \in R^{i}$. If $\left( a_{1},a_{2}\right)
\in R^{i}$, then $\left( a_{0},a_{2}\right) \in R^{i}$ because $R^{i}$ is
transitive. Suppose $\left( a_{1},a_{2}\right) \notin R^{i}$. Then, $%
(a_{2},a_{1})\notin P\left( R^{i}\right) $ implies $(a_{2},a_{1})\notin
R^{i} $. Moreover, toward a contradiction, we assume $\left(
a_{0},a_{2}\right) \notin R^{i}$. Since $R^{i}$ is negatively transitive, $%
\left( a_{0},a_{2}\right) \notin R^{i}$ and $(a_{2},a_{1})\notin R^{i}$
imply $\left( a_{0},a_{1}\right) \notin R^{i}$, which is a contradiction.
Therefore, $\left( a_{0},a_{2}\right) \in R^{i}$. Likewise, we can show $%
\left( a_{0},a_{K}\right) \in R^{i}$ and thus $\left( a_{K},a_{0}\right)
\notin R_{\alpha }$.

\section*{Appendix B: Proofs of Main Results}

\subsection*{Proof of Proposition 1}

The only-if-part of the first result is trivial. Moreover, the if-part of
the first result is trivial from the second one. Thus, we only show the
second result.

First, by Lemma 2, if $R$ is P-acyclic, the maximal set is not empty for any
step and thus each algorithm in the SMO class is well-defined.

Suppose that $\hat{R}:a_{1},a_{2},\cdots ,a_{\left\vert A\right\vert }$ is a
linear order obtained by the SMO algorithm class. Then, for all $t$ and $%
t^{\prime }$ such that $1\leq t<t^{\prime }\leq \left\vert A\right\vert $, $%
\left( a_{t},a_{t^{\prime }}\right) \notin P\left( R\right) $. Therefore, $%
\hat{R}\in \mathcal{C}\left( R\right) $. Second, suppose $\hat{R}\left(
:a_{1},a_{2},\cdots ,a_{\left\vert A\right\vert }\right) \in \mathcal{R}%
\left( R\right) $. Suppose not; that is, $\hat{R}$ is not obtained by any
algorithm within the SMO class with $R$. Then, we can let $t$ be the
smallest integer such that $a_{t}\notin M\left( A\setminus \left\{
a_{1},\cdots ,a_{t-1}\right\} ,R\right) $; that is, there is $t^{\prime }>t$
such that $\left( a_{t^{\prime }},a_{t}\right) \in P\left( R\right) $, which
contradicts $\hat{R}\in \mathcal{R}\left( R\right) $.

\subsection*{Proof of Proposition 2}

We show the first result. First, since $T\left( R_{\alpha }\right) $ is
transitive, it is P-acyclic (by Lemma 1). By Proposition 1, $\emptyset \neq 
\mathcal{R}\left( T\left( R_{\alpha }\right) \right) $ and thus, we can let $%
\hat{R}\in \mathcal{R}\left( T\left( R_{\alpha }\right) \right) $. By the
definition of respecting linear orders, for any $\left( a,b\right) \in
P\left( T\left( R_{\alpha }\right) \right) ,$ $\left( a,b\right) \in \hat{R}$%
. Then, since $P\left( S\left( R_{\alpha }\right) \right) \subseteq P\left(
T\left( R_{\alpha }\right) \right) $ (by Lemma 5), for any $\left(
a,b\right) \in P\left( S\left( R_{\alpha }\right) \right) $, $\left(
a,b\right) \in \hat{R}$. Therefore, $\hat{R}\in \mathcal{R}\left( S\left(
R_{\alpha }\right) \right) $.

We show the second result. Suppose $\mathcal{R}\left( R_{\alpha }\right)
\neq \emptyset $. By Proposition 1, $R_{\alpha }$ is $P$-acyclic. By the
first result of Lemma 7, $R_{\alpha }$ is asymmetric. By the third result of
Lemma 1, $R_{\alpha }$ is Suzumura-consistent and thus $S\left( R_{\alpha
}\right) =R_{\alpha }$. We have $\mathcal{R}\left( S\left( R_{\alpha
}\right) \right) =\mathcal{R}\left( R_{\alpha }\right) $.

By the first result, showing $\mathcal{R}\left( S\left( R_{\alpha }\right)
\right) \subseteq \mathcal{R}\left( T\left( R_{\alpha }\right) \right) $ is
sufficient. Suppose not; that is, there is $R\in \mathcal{R}\left( S\left(
R_{\alpha }\right) \right) \setminus \mathcal{R}\left( T\left( R_{\alpha
}\right) \right) $. Then, there is $\left( a,b\right) \in R$ such that $%
\left( b,a\right) \in P\left( T\left( R_{\alpha }\right) \right) \setminus
P\left( S\left( R_{\alpha }\right) \right) $. Then, $\left( a,b\right)
\notin T\left( R_{\alpha }\right) $ implies $\left( a,b\right) \notin
S\left( R_{\alpha }\right) $. Therefore, $\left( b,a\right) \in T\left(
R_{\alpha }\right) $; that is, $\left( a_{1},a_{2},\cdots ,a_{m}\right) \in 
\mathcal{B}\left( \left[ b,a\right] ,R_{\alpha }\right) $, where $a_{1}=b$
and $a_{m}=a$. If $\left( a_{l},a_{k}\right) \in T\left( R_{\alpha }\right) $
for some $m\geq l>k\geq 1$, then there is a $P$-cycle, because $R_{\alpha }$
is asymmetric (by the first result of Lemma 7). Thus, $\left(
a_{l},a_{k}\right) \notin T\left( R_{\alpha }\right) $ and moreover $\left(
a_{l},a_{k}\right) \notin S\left( R_{\alpha }\right) $ for all $m\geq
l>k\geq 1$. Therefore, $\left( a_{l},a_{l+1}\right) \in P\left( S\left(
R_{\alpha }\right) \right) $ and $\left( a_{l},a_{l+1}\right) \in R$ for all 
$l=1,2,\cdots ,m-1$. Since $R$ is a linear order, $\left( b,a\right) \in R$
contradicts $\left( a,b\right) \in R$.

\subsection*{Proof of Proposition 3}

First, let $1>\alpha \geq \beta \geq 0$. By the third\ result of Lemma 7, if 
$R_{\beta }$ is P-acyclic, then $R_{\alpha }$ is also P-acyclic, and if $%
R_{\alpha }$ has a P-cycle, then $R_{\beta }$ also has a P-cycle. Therefore, 
$R_{\alpha }$ is P-acyclic if and only if $\alpha \geq \alpha ^{\ast }$. We
have $\bigcap\nolimits_{\alpha :\text{ }R_{\alpha }\in \mathcal{P}_{^{AC}}}%
\mathcal{R}\left( R_{\alpha }\right) =\bigcap\nolimits_{\alpha \geq \alpha
^{\ast }}\mathcal{R}\left( R_{\alpha }\right) $.

Since $R_{\alpha }$ is asymmetric (by the first result of Lemma 7), $P\left(
R_{\alpha }\right) =R_{\alpha }$. Moreover, since $R_{\alpha ^{\ast
}}\subseteq R_{\alpha }$ for all $\alpha \in \left[ \alpha ^{\ast },1\right) 
$ (by the second result of Lemma 7)$,$ $\bigcup_{\alpha \geq \alpha ^{\ast
}}P\left( R_{\alpha }\right) =\bigcup_{\alpha \geq \alpha ^{\ast }}R_{\alpha
}=R_{\alpha ^{\ast }}$. Thus, by Lemma 4, we have $\bigcap\nolimits_{\alpha :%
\text{ }R_{\alpha }\in \mathcal{P}_{^{AC}}}\mathcal{R}\left( R_{\alpha
}\right) =\mathcal{R}\left( R_{\alpha ^{\ast }}\right) $. Finally, by
Proposition 1 and Lemma 9, $\mathcal{R}\left( R_{\alpha ^{\ast }}\right)
\neq \emptyset .$

\subsection*{Proof of Theorem 1}

By Lemma 6 and the second result of Lemma 7, $S\left( R_{\alpha }\right)
\subseteq S\left( R_{1/2}\right) $ for all $\alpha \in \left[ 1/2,1\right) $%
. Since $R_{1/2}$ is finite, we can let be finite binary relations $R\left(
1\right) ,R\left( 2\right) ,\cdots ,R\left( I\right) $ such that for each $%
\alpha \in \left[ 1/2,1\right) $, $S\left( R_{\alpha }\right) =R\left(
i\right) $ for some $i=1,\cdots ,I,$ and $R\left( 1\right) \subseteq R\left(
2\right) \subseteq \cdots \subseteq R\left( I\right) $, because of Lemma 6
and the second result of Lemma 7. Moreover, since $S\left( R_{\alpha
}\right) $ is P-acyclic for all $\alpha \in \left[ 1/2,1\right) $ (by the
second result of Lemma 1), $R\left( i\right) $ is P-acyclic for all $%
i=1,\cdots ,I$. By Lemma 4, we have the equality of (\ref{e}).

\subsection*{Proof of Proposition 4}

The\textbf{\ }asymmetry of $R^{\pi }$ is trivial. First, we show the
completeness of $R^{\pi }$. Suppose not; that is, $\left( a,b\right) \notin
R^{\pi }$ and $\left( b,a\right) \notin R^{\pi }$. Then, $\mathcal{B}\left( %
\left[ b,a\right] ,R_{\pi \left( a,b\right) -1}^{\pi }\right) \neq \emptyset 
$ and $\mathcal{B}\left( \left[ a,b\right] ,R_{\pi \left( b,a\right)
-1}^{\pi }\right) \neq \emptyset $.

In Step 1, there is no path from $a$ to $b$ or from $b$ to $a$ for $%
R_{1}^{\pi }=\emptyset $. On the other hand, since $\left( a,b\right) \notin
R^{\pi }$ and $\left( b,a\right) \notin R^{\pi }$, there are at least two
paths from $a$ to $b$ and from $b$ to $a$ for $R_{\left\vert A\right\vert
^{2}}^{\pi }=R^{\pi }$ in the terminal step (Step $\left\vert A\right\vert
^{2}$). Thus, they are made in the process. Let $\left( a_{1},a_{2},\cdots
,a_{m}\right) $ be a path for $R^{\pi }$ such that it is firstly made among
the paths from $a$ to $b$ and from $b$ to $a$ in the process. Without loss
of generality, suppose $\left( a_{1},a_{2},\cdots ,a_{m}\right) $ is from $%
a\left( =a_{1}\right) $ to $b\left( =a_{m}\right) $. Then, consider \textbf{%
Step} $z$ such that a path $\left( b_{1},b_{2},\cdots ,b_{m^{\prime
}}\right) $ from $b\left( =b_{1}\right) $ to $a\left( =b_{m^{\prime
}}\right) $ for $R^{\pi }$ is newly made. Then, we can let $\pi \left(
b_{l},b_{l+1}\right) =z$ for some $l$; that is, $R_{z}^{\pi }=R_{z-1}^{\pi
}\cup \left\{ \left( b_{l},b_{l+1}\right) \right\} $ and $\left(
b_{k},b_{k+1}\right) \in R_{z-1}^{\pi }$ for all $k\neq l$. Then, since $%
a_{1}=b_{m^{\prime }}=a$ and $b_{1}=a_{m}=b,$ 
\begin{equation*}
\left( b_{l+1},b_{l+2},\cdots ,b_{m^{\prime }}\left( =a_{1}\right)
,a_{2},\cdots ,a_{m}\left( =b_{1}\right) ,\cdots ,b_{l}\right)
\end{equation*}%
is a path for $R_{z-1}^{\pi }$ from $b_{l+1}$ to $b_{l}$; that is, $\mathcal{%
B}\left( \left[ b_{l+1},b_{l}\right] ,R_{z-1}^{\pi }\right) \neq \emptyset $%
. However, this contradicts $R_{z}^{\pi }=R_{z-1}^{\pi }\cup \left\{ \left(
b_{l},b_{l+1}\right) \right\} $. Therefore, $R^{\pi }$ is complete.

Second, we show transitivity of $R^{\pi }$. Suppose $\left( a,b\right)
,\left( b,c\right) \in R^{\pi }$. Then, showing $\left( c,a\right) \notin
R^{\pi }$ is sufficient, because $R^{\pi }$ is complete. First, suppose $\pi
\left( c,a\right) >\max \left\{ \pi \left( a,b\right) ,\pi \left( b,c\right)
\right\} $. Then, since $\left( a,b,c\right) \in \mathcal{B}\left( \left[ a,c%
\right] ,R_{\pi \left( c,a\right) -1}^{\pi }\right) \neq \emptyset $, $%
\left( c,a\right) \notin R^{\pi }$. Second, suppose $\pi \left( b,c\right)
>\max \left\{ \pi \left( a,b\right) ,\pi \left( c,a\right) \right\} $. Then,
since $\mathcal{B}\left( \left[ c,b\right] ,R_{\pi \left( b,c\right)
-1}^{\pi }\right) =\emptyset $, $\left( c,a,b\right) \notin \mathcal{B}%
\left( \left[ c,b\right] ,R_{\pi \left( b,c\right) -1}^{\pi }\right) $.
Thus, since $\left( a,b\right) \in R_{\pi \left( b,c\right) -1}^{\pi }$, $%
\left( c,a\right) \notin $ $R_{\pi \left( b,c\right) -1}^{\pi }$ and $\left(
c,a\right) \notin $ $R^{\pi }$. Third, suppose $\pi \left( a,b\right) >\max
\left\{ \pi \left( b,c\right) ,\pi \left( c,a\right) \right\} $. Then, since 
$\mathcal{B}\left( \left[ b,a\right] ,R_{\pi \left( a,b\right) -1}^{\pi
}\right) =\emptyset $ and $\left( b,c\right) \in R_{\pi \left( a,b\right)
-1}^{\pi }$, $\left( c,a\right) \notin R_{\pi \left( a,b\right) -1}^{\pi }$
and hence $\left( c,a\right) \notin $ $R^{\pi }$. Therefore, $R^{\pi }$ is
transitive.

Finally, we show (\ref{f}). Since we have already shown that $R^{\pi }$ is a
linear order, showing $\bigcup_{\alpha \in \left[ 1/2,1\right) \ }P\left(
S\left( R_{\alpha }\right) \right) \subseteq R^{\pi }$ for any $\pi \in \Pi $
is sufficient. Let $\left( a,b\right) \in P\left( S\left( R_{\alpha }\right)
\right) $ for some $\alpha \in \left[ 1/2,1\right) $. By Lemma 5, $\left(
a,b\right) \in P\left( R_{\alpha }\right) $ and thus $\left( a,b\right) \in
R_{\alpha }$ and $\left( b,a\right) \notin R_{\alpha }$. Suppose not; that
is, $\left( a,b\right) \notin R^{\pi }$ for some $\pi \in \Pi $. Let $z=\pi
\left( a,b\right) $. Then, we can let $\left( a_{1},a_{2},\cdots
,a_{m}\right) \in \mathcal{B}\left( \left[ b,a\right] ,R_{z-1}^{\pi }\right)
\neq \emptyset $, where $a_{1}=b$ and $a_{m}=a$. By construction, $\left(
a_{l},a_{l+1}\right) \in R_{z-1}^{\pi }$ for all $l=1,\cdots ,m-1$ implies $%
\pi \left( a_{l},a_{l+1}\right) >\pi \left( a,b\right) $ and thus $\Phi %
\left[ a_{l},a_{l+1}\right] \geq \Phi \lbrack a,b]$ for all $l=1,\cdots ,m-1$%
. Since $\left( a,b\right) \in R_{\alpha },$ $\Phi \left[ a_{l},a_{l+1}%
\right] \geq \Phi \lbrack a,b]>\alpha $. Thus, $\left( a_{l},a_{l+1}\right)
\in R_{\alpha }$ for all $l=1,\cdots m-1$. Hence $\left( a_{1},a_{2},\cdots
,a_{m}\right) \in \mathcal{B}\left( \left[ b,a\right] ,R_{\alpha }\right) $.
Then, since $\left( a,b\right) \in R_{\alpha }$, $\left( b,a\right) \in
S\left( R_{\alpha }\right) $ but this contradicts $\left( a,b\right) \in
P\left( S\left( R_{\alpha }\right) \right) $.

\subsection*{Proof of Proposition 5}

The first fact is trivial from the first result of Proposition 2.

We show the second result. Suppose that $R^{i}$ is a weak order for all $%
i\in I$. It is sufficient to show $\mathcal{S\subseteq }\bigcap\nolimits_{%
\alpha :\text{ }R_{\alpha }\in \mathcal{P}_{^{AC}}}\mathcal{R}\left(
R_{\alpha }\right) $. By Proposition 3, $\alpha ^{\ast }$ is well-defined
and $\mathcal{R}\left( R_{\alpha ^{\ast }}\right) \neq \emptyset $. By the
third result of Lemma 1 and the frist result of Lemma 7, $S\left( R_{\alpha
^{\ast }}\right) =R_{\alpha ^{\ast }}$. Moreover, by Proposition 3, $%
\bigcap\nolimits_{\alpha :\text{ }R_{\alpha }\in \mathcal{P}_{^{AC}}}%
\mathcal{R}\left( R_{\alpha }\right) =\mathcal{R}\left( R_{\alpha ^{\ast
}}\right) $. Now, let $\hat{R}\in \mathcal{S}$. Then, $\left( a,b\right) \in
P\left( S\left( R_{\alpha ^{\ast }}\right) \right) =P\left( R_{\alpha ^{\ast
}}\right) $ implies $\left( a,b\right) \in \hat{R}$. Thus, by the definition
of $\mathcal{R}\left( R_{\alpha ^{\ast }}\right) $, $\hat{R}\in \mathcal{R}%
\left( R_{\alpha ^{\ast }}\right) =\bigcap\nolimits_{\alpha :\text{ }%
R_{\alpha }\in \mathcal{P}_{^{AC}}}\mathcal{R}\left( R_{\alpha }\right) $.

\subsection*{Proof of Proposition 6}

We first show $R^{Sc}\subseteq \bigcup_{\alpha \in \left[ 1/2,1\right)
}P\left( T\left( R_{\alpha }\right) \right) $. Let $\left( a,b\right) \in
R^{Sc}$; that is, $B[a,b]>B[b,a]$.

First, suppose $B[b,a]<1/2$. By the definition of $B[b,a]$, $B[b,a]=0$ and $%
\mathcal{B}\left[ \left( b,a\right) ,R_{1/2}\right] =\emptyset $. Thus, $%
\left( b,a\right) \notin T\left( R_{1/2}\right) $. By Lemma 6 and the second
result of Lemma 7, $\left( b,a\right) \notin T\left( R_{\alpha }\right) $
for all $\alpha \in \left[ 1/2,1\right) $ and thus $\left( b,a\right) \notin
\bigcup_{\alpha \in \left[ 1/2,1\right) }P\left( T\left( R_{\alpha }\right)
\right) $. Since $B[a,b]>B[b,a]=0$, $\mathcal{B}\left[ \left( a,b\right)
,R_{1/2}\right] \neq \emptyset $ and thus $\left( a,b\right) \in T\left(
R_{1/2}\right) $. Therefore, in this case, 
\begin{equation*}
\left( a,b\right) \in P\left( T\left( R_{1/2}\right) \right) \subseteq
\bigcup_{\alpha \in \left[ 1/2,1\right) }P\left( T\left( R_{\alpha }\right)
\right) .
\end{equation*}

Second, suppose $B[b,a]\in \left[ 1/2,1\right) $. By Lemma 8, $\mathcal{B}%
\left[ \left( b,a\right) ,R_{B[b,a]}\right] =\emptyset $ and $\mathcal{B}%
\left[ \left( a,b\right) ,R_{B[b,a]}\right] \neq \emptyset $. That is, $%
\left( a,b\right) \in T\left( R_{B[b,a]}\right) $ and $\left( b,a\right)
\notin T\left( R_{B[b,a]}\right) $. Since $B[b,a]\in \left[ 1/2,1\right) $,
we have 
\begin{equation*}
\left( a,b\right) \in P\left( T\left( R_{B[b,a]}\right) \right) \subseteq
\bigcup_{\alpha \in \left[ 1/2,1\right) }P\left( T\left( R_{\alpha }\right)
\right) .
\end{equation*}%
Therefore, $R^{Sc}\subseteq \bigcup_{\alpha \in \left[ 1/2,1\right) }P\left(
T\left( R_{\alpha }\right) \right) $.

Next, we show $R^{Sc}\supseteq \bigcup_{\alpha \in \left[ 1/2,1\right)
}P\left( T\left( R_{\alpha }\right) \right) $. Fix some $\alpha \in \left[
1/2,1\right) $ and let $\left( a,b\right) \in P\left( T\left( R_{\alpha
}\right) \right) $; that is, $\left( a,b\right) \in T\left( R_{\alpha
}\right) $ and $\left( b,a\right) \notin T\left( R_{\alpha }\right) $.
Toward a contradiction, suppose $\left( a,b\right) \notin R^{Sc}$. Then, $%
B[a,b]\leq B[b,a]$. First, suppose $B[a,b]<B[b,a]$. Then $\left( b,a\right)
\in R^{Sc}$. Since $R^{Sc}\subseteq \bigcup_{\alpha \in \left[ 1/2,1\right)
}P\left( T\left( R_{\alpha }\right) \right) $, $\left( b,a\right) \in
P\left( T\left( R_{\alpha ^{\prime }}\right) \right) $; that is, $\left(
b,a\right) \in T\left( R_{\alpha ^{\prime }}\right) $ and $\left( a,b\right)
\notin T\left( R_{\alpha ^{\prime }}\right) ,$ for some $\alpha ^{\prime
}\in \left[ 1/2,1\right) $. First, suppose $\alpha ^{\prime }<\alpha $.
Then, by Lemma 6 and the second result of Lemma 7, $T\left( R_{\alpha
}\right) \subseteq T\left( R_{\alpha ^{\prime }}\right) ,$ which contradicts 
$\left( a,b\right) \in T\left( R_{\alpha }\right) $ and $\left( a,b\right)
\notin T\left( R_{\alpha ^{\prime }}\right) $. Second, suppose $\alpha
^{\prime }\geq \alpha $. Then, $T\left( R_{\alpha ^{\prime }}\right)
\subseteq T\left( R_{\alpha }\right) ,$ which contradicts $\left( b,a\right)
\notin T\left( R_{\alpha }\right) $ and $\left( b,a\right) \in T\left(
R_{\alpha ^{\prime }}\right) $. Therefore, if $B[a,b]<B[b,a]$, then $\left(
a,b\right) \in R^{Sc}$ and $R^{Sc}\supseteq \bigcup_{\alpha \in \left[
1/2,1\right) }P\left( T\left( R_{\alpha }\right) \right) $.

Second, suppose $B[a,b]=B[b,a]$. Then, since $\left( a,b\right) \in T\left(
R_{\alpha }\right) $, $\mathcal{B}\left[ \left( a,b\right) ,R_{\alpha }%
\right] \neq \emptyset $. By Lemma 8, $\alpha <B[a,b]=B[b,a]$. Then, by the
definition of $B[b,a]$, there is a sequence of alternatives$\ \left(
a_{1},a_{2},\cdots ,a_{m}\right) $ such that $a_{1},a_{2},\cdots ,a_{m}$ are
distinct, $a_{1}=b$, $a_{m}=a$, and $\Phi \left[ a_{j},a_{j+1}\right] \geq
B[b,a]>\alpha $ for all $j=1,\cdots ,m-1$. Then, $\left( a_{1},a_{2},\cdots
,a_{m}\right) \in \mathcal{B}\left[ \left( b,a\right) ,R_{\alpha }\right] $,
which contradicts $\left( b,a\right) \notin T\left( R_{\alpha }\right) $.

\subsection*{Proof of Proposition 7}

Fix $\mathbf{A}=\left\{ A_{1},\cdots ,A_{X}\right\} \in \mathcal{A}$. First,
let $a$ and $a^{\prime }$ be such that $a\in A_{x}$ and $a^{\prime }\in
A_{x^{\prime }}$ with $x<x^{\prime }$. We show $\left( a,a^{\prime }\right)
\in P\left( S\left( R_{1/2}\right) \right) $. Suppose not; that is, $\left(
a,a^{\prime }\right) \notin P\left( S\left( R_{1/2}\right) \right) $. Then,
since $N\left[ a,a^{\prime }\right] >N\left[ a^{\prime },a\right] $, $\left(
a,a^{\prime }\right) \in R_{1/2}$ (by the fourth result of Lemma 7) and thus 
$\left( a,a^{\prime }\right) \in S\left( R_{1/2}\right) $. Therefore, $%
\left( a,a^{\prime }\right) \notin P\left( S\left( R_{1/2}\right) \right) $
implies $\left( a^{\prime },a\right) \in S\left( R_{1/2}\right) $. Since $%
R_{1/2}$ is asymmetric (by the first result of Lemma 8), there is a path $%
\left( a_{1},a_{2},\cdots ,a_{m}\right) \in \mathcal{B}\left[ \left(
a^{\prime },a\right) ,R_{1/2}\right] $, where $a_{1}=a^{\prime }$, and $%
a_{m}=a$ and $\left( a_{l},a_{l+1}\right) \in R_{1/2}$ for all $l=1,2,\cdots
,m-1$. By the fourth result of Lemma 7, $\left( a_{l},a_{l+1}\right) \in
R_{1/2}$ implies $N\left[ a_{l},a_{l+1}\right] >N\left[ a_{l+1},a_{l}\right] 
$ for all $l=1,2,\cdots ,m-1$. Since $a_{1}=a^{\prime }\in A_{x^{\prime }}$
and $N\left[ a_{1},a_{2}\right] >N\left[ a_{2},a_{1}\right] $, 
\begin{equation*}
a_{2}\in A_{x^{\prime }}\cup \cdots \cup A_{X}\text{.}
\end{equation*}%
Moreover, since $N\left[ a_{2},a_{3}\right] >N\left[ a_{3},a_{2}\right] $, $%
a_{3}$ also in the set. Similarly, we have 
\begin{equation*}
a_{2},a_{3},\cdots ,a_{m}\in A_{x^{\prime }}\cup \cdots \cup A_{X}\text{,}
\end{equation*}%
which contradict to $a_{m}=a\in A_{x}$ and $a^{\prime }\in A_{x^{\prime }}$
with $x<x^{\prime }$.

\subsection*{Proof of Proposition 8}

We fix $\alpha \in \left[ f\left( \left\vert I\right\vert -1,1\right)
,1\right) $. Suppose that $\left( b,a\right) \notin P\left( R^{i}\right) $
for all $i\in I$ and $\left( a,b\right) \in P\left( R^{i}\right) $ for some $%
i\in I$. We show $\left( a,b\right) \in \hat{R}$. Since $R^{i}$ is a linear
order, $\left( b,a\right) \notin P\left( R^{i}\right) $ implies that $\left(
a,b\right) \in P\left( R^{i}\right) $. Therefore, $N\left[ b,a\right] =0$
and $N\left[ a,b\right] =\left\vert I\right\vert $. Then, $\Phi \left[ a,b%
\right] =f\left( \left\vert I\right\vert ,0\right) =1$ and $\Phi \left[ b,a%
\right] =f\left( 0,\left\vert I\right\vert \right) <1/2$. Thus, $\left(
a,b\right) \in R_{\alpha }$ and $\left( b,a\right) \notin R_{\alpha }$. By
Lemma 9, $R_{\alpha }$ is P-acyclic. By the third result of Lemma 1, $%
R_{\alpha }$ is Suzumura-consistent and thus $S\left( R_{\alpha }\right)
=R_{\alpha }$. Then, $\left( a,b\right) \in P\left( S\left( R_{\alpha
}\right) \right) =P\left( R_{\alpha }\right) $ and hence $\left( a,b\right)
\in \hat{R}$.

\subsection*{Proof of Proposition 9}

We fix $\alpha \in \left[ f\left( \left\vert I\right\vert -1,1\right)
,1\right) $. Then, we have $f\left( m,n\right) \leq \alpha $ and $f\left(
n,0\right) =1>\alpha $ for all $m\in \left[ 0,\left\vert I\right\vert -1%
\right] $ and all $n>0$. Then, $\left( a,b\right) \in R_{\alpha }$ if and
only if $\left( b,a\right) \notin R^{i}$ for all $i\in I$ and $\left(
a,b\right) \in R^{i}$ for some $i\in I$.

By Lemma 9, $R_{\alpha }$ is P-acyclic. By Proposition 1, $\mathcal{R}\left(
R_{\alpha }\right) \neq \emptyset $. By Proposition 2, $\mathcal{R}\left(
R_{\alpha }\right) =\mathcal{R}\left( S\left( R_{\alpha }\right) \right) $.
Thus, for any $\hat{R}\in \mathcal{R}\left( S\left( R_{\alpha }\right)
\right) ,$ $\hat{R}\in \mathcal{R}\left( R_{\alpha }\right) $. Now, suppose
that there are $a,b\in A$ such that $\left( b,a\right) \notin P\left(
R^{i}\right) $ for all $i\in I$ and $\left( a,b\right) \in P\left(
R^{i}\right) $ for some $i\in I$. Then, since $N\left[ b,a\right] =0$ and $N%
\left[ a,b\right] >0$, $\left( a,b\right) \in P\left( R_{\alpha }\right) $.
Hence $\left( a,b\right) \in \hat{R}$.

\subsection*{Proof of Proposition 10}

First, we show the if-part. Suppose $\Phi \left[ a,b\right] >1/2$ and $\Phi %
\left[ a,b\right] >B[b,a]$. Then, for any $\alpha \in \left[ 1/2,\Phi \left[
a,b\right] \right) $, $(a,b)\in R_{\alpha }\subseteq S\left( R_{\alpha
}\right) $. First, we show that $(b,a)\notin S\left( R_{\alpha }\right) $
for some $\alpha \in \left[ 1/2,\Phi \left[ a,b\right] \right] $. Since $%
R_{\alpha }$ is asymmetric (the first result of Lemma 7), $(b,a)\notin
R_{\alpha }$ for any $\alpha \in \left[ 1/2,\Phi \left[ a,b\right] \right) $%
. Moreover, by the second result of Lemma 7, $(b,a)\notin R_{\alpha }$ for
all $\alpha \in \left[ 1/2,1\right) $.

If $B[b,a]<1/2$, then $(b,a)\notin T\left( R_{\alpha }\right) $ and
therefore, $(b,a)\notin S\left( R_{\alpha }\right) $ for all $\alpha \in %
\left[ 1/2,1\right) $. Thus, we assume $B[b,a]\geq 1/2$. In this case, $%
(b,a)\in T\left( R_{\alpha }\right) $ if and only if $\alpha \in \left[
1/2,B[b,a]\right) $. Therefore, for $\alpha \in \left[ B[b,a],\Phi \left[ a,b%
\right] \right) $, $(b,a)\notin T\left( R_{\alpha }\right) .$ Since $S\left(
R_{\alpha }\right) \subseteq T\left( R_{\alpha }\right) $, $\left(
b,a\right) \notin S\left( R_{\alpha }\right) $, for $\alpha \in \left[
B[b,a],\Phi \left[ a,b\right] \right) $. Therefore, since $\left( a,b\right)
\in S\left( R_{\alpha }\right) $ for any $\alpha \in \left[ 1/2,\Phi \left[
a,b\right] \right] $, $\left( a,b\right) \in P\left( S\left( R_{\alpha
}\right) \right) $ for $\alpha \in \left[ B[b,a],\Phi \left[ a,b\right]
\right) $.

Second, we show the only-if-part. Suppose $\left( a,b\right) \in
\bigcup_{\alpha \in \left[ 1/2,1\right) }P\left( S\left( R_{\alpha }\right)
\right) $. Then, we can fix $\alpha \in \left[ 1/2,1\right) $ such that $%
\left( a,b\right) \in S\left( R_{\alpha }\right) $ and $\left( b,a\right)
\notin S\left( R_{\alpha }\right) $. Then, by Lemma 5, $\left( a,b\right)
\in P\left( S\left( R_{\alpha }\right) \right) $ implies $\left( a,b\right)
\in P\left( R_{\alpha }\right) $; that is, $\left( a,b\right) \in R_{\alpha
} $ and $\left( b,a\right) \notin R_{\alpha }$. Thus, $\Phi \left[ a,b\right]
>1/2$. Next, since $S\left( R_{\alpha }\right) \subseteq T\left( R_{\alpha
}\right) $ and $\left( b,a\right) \notin S\left( R_{\alpha }\right) $, $%
\left( b,a\right) \notin T\left( R_{\alpha }\right) $. Then, $\alpha $ must
be more than $B[b,a]$ and less than or equal to $\Phi \left[ a,b\right] $.
Hence $\Phi \left[ a,b\right] >B[b,a]$.

\section*{Appendix C: Example}

Here, we provide an example where the linear orders obtained by the Schulze
method, the ranked pairs method, the Kemeny-Young method, and the Borda
method differ.

Let $A=\left\{ a,b,c,d,e\right\} $ and $I=\left\{ 1,\cdots ,47\right\} $.
All individuals have a linear order on $A$ and%
\begin{align*}
R^{i}& :\text{ }a,c,b,e,d\text{ for }i=1,\cdots ,5, \\
& :\text{ }a,d,e,c,b\text{ for }i=6,\cdots ,10, \\
& :\text{ }c,a,b,e,d\text{ for }i=11,12,13 \\
& :\text{ }c,a,e,b,d\text{ for }i=14,\cdots ,20, \\
& :\text{ }b,c,d,a,e\text{ for }i=21 \\
& :\text{ }b,e,d,a,c\text{ for }i=22,\cdots ,29, \\
& :\text{ }c,b,a,d,e\text{ for }i=30,31, \\
& :\text{ }e,b,a,d,c\text{ for }i=32,\cdots ,39, \\
& :\text{ }d,c,b,a,e\text{ for }i=40 \\
& :\text{ }d,c,e,b,a\text{ for }i=41,\cdots ,44, \\
& :\text{ }d,e,c,b,a\text{ for }i=45,46,47.
\end{align*}%
Then, the following table summarizes the voting results, where the number of
a cell represents that of individuals who prefers the row alternative to the
column one; that is, for example, $N\left[ b,a\right] =27$, $N\left[ e,d%
\right] =31$ and $N\left[ c,e\right] =23$.

\begin{center}
\begin{tabular}{|l|l|l|l|l|l|}
\hline
& $a$ & $b$ & $c$ & $d$ & $e$ \\ \hline
$a$ &  & 20 & 26 & 30 & 24 \\ \hline
$b$ & 27 &  & 17 & 34 & 20 \\ \hline
$c$ & 21 & 30 &  & 18 & 23 \\ \hline
$d$ & 17 & 13 & 29 &  & 16 \\ \hline
$e$ & 23 & 27 & 24 & 31 &  \\ \hline
\end{tabular}
\end{center}

Then $R_{1/2}$ is represented in \textbf{Figure 1}; that is, there is an
arrow from from $j$ to $k$ exists if $\left( j,k\right) \in R_{1/2}$ or
equivalently if 24 or more individuals prefer $j$ to $k$, where the number
written nearby the arrow from $j$ to $k$ represents the number of
individuals who prefer $j$ to $k$.

\begin{figure}[tbph]
\begin{center}
\includegraphics[width=110mm]{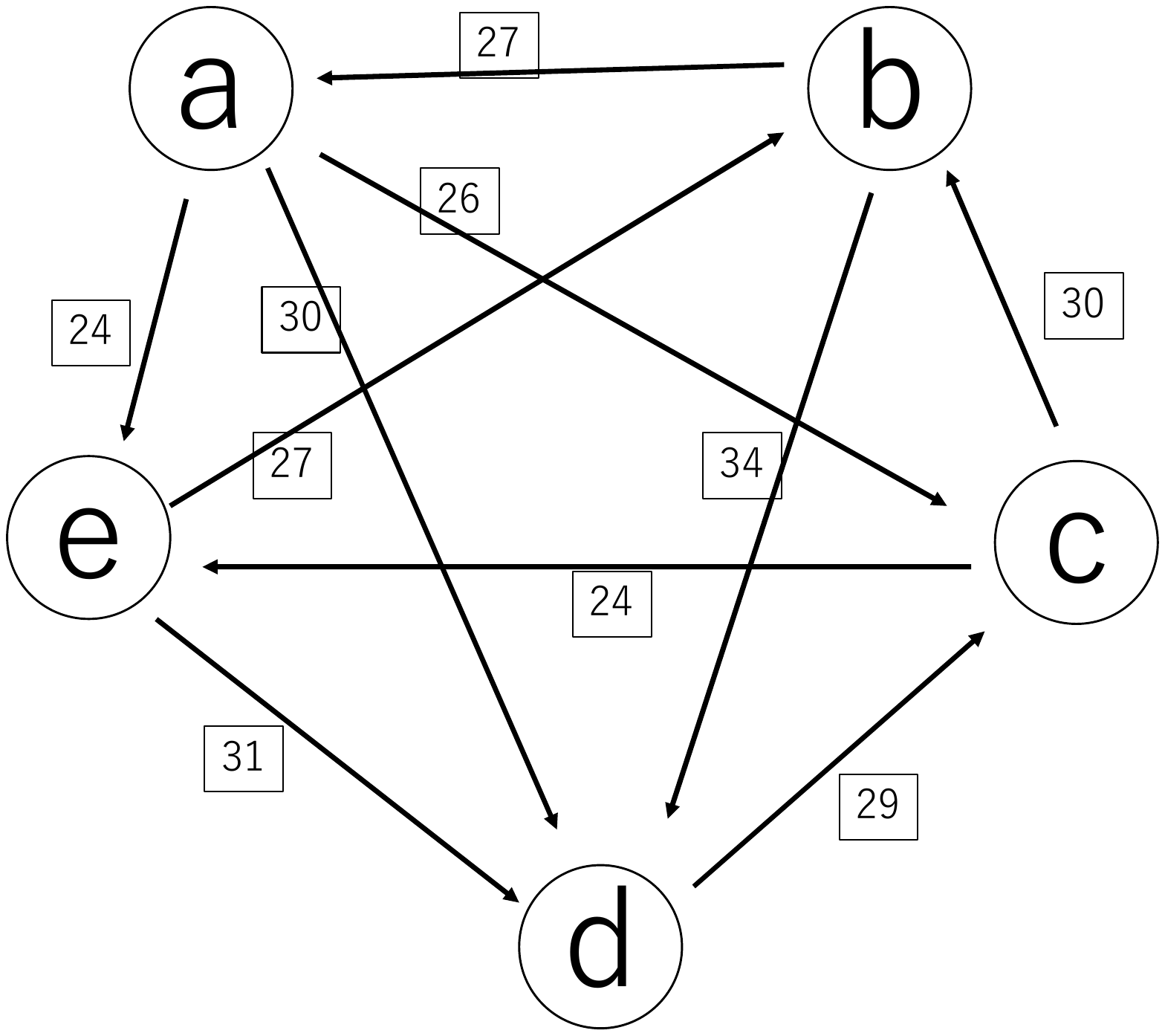}
\end{center}
\caption{$R_{1/2}$ of this example}
\end{figure}

When we adopt (\ref{b}) or (\ref{b1}), $R_{\alpha },$ $S\left( R_{\alpha
}\right) $ and $T\left( R_{\alpha }\right) $ are as follows. First, for $%
\alpha \in \left[ 1/2,24/47\right) ,$%
\begin{eqnarray*}
R_{\alpha } &=&\left\{ 
\begin{array}{c}
\left( a,c\right) ,\left( a,d\right) ,\left( a,e\right) ,\left( b,a\right)
,\left( b,d\right) , \\ 
\left( c,b\right) ,\left( c,e\right) ,\left( d,c\right) ,\left( e,b\right)
,\left( e,d\right)%
\end{array}%
\right\} , \\
S\left( R_{\alpha }\right) &=&\left\{ \left( a,d\right) ,\left( b,d\right)
\right\} =T\left( R_{\alpha }\right) \text{ }.
\end{eqnarray*}%
Second, for $\alpha \in \left[ 24/47,26/47\right) ,$%
\begin{eqnarray*}
R_{\alpha } &=&R_{1/2}\setminus \left\{ \left( a,e\right) ,\left( c,e\right)
\right\} ,\text{ } \\
P\left( S\left( R_{\alpha }\right) \right) &=&\left\{ \left( e,b\right)
,\left( e,d\right) \right\} , \\
P\left( T\left( R_{\alpha }\right) \right) &=&P\left( S\left( R_{\alpha
}\right) \right) \cup \left\{ \left( e,a\right) ,\left( e,c\right) \right\} .
\end{eqnarray*}%
Third, for $\alpha \in \left[ 26/47,27/47\right) ,$%
\begin{eqnarray*}
R_{\alpha } &=&R_{24/47}\setminus \left\{ \left( a,c\right) \right\} ,\text{ 
} \\
P\left( S\left( R_{\alpha }\right) \right) &=&\left\{ \left( e,b\right)
,\left( e,d\right) \right\} , \\
P\left( T\left( R_{\alpha }\right) \right) &=&P\left( S\left( R_{\alpha
}\right) \right) \cup \left\{ \left( e,a\right) ,\left( e,c\right) \right\} .
\end{eqnarray*}%
Fourth, for $\alpha \in \left[ 27/47,29/47\right) ,$%
\begin{eqnarray*}
R_{\alpha } &=&R_{26/47}\setminus \left\{ \left( b,a\right) ,\left(
e,b\right) \right\} \\
P\left( S\left( R_{\alpha }\right) \right) &=&\left\{ \left( a,d\right)
,\left( e,d\right) \right\} , \\
P\left( T\left( R_{\alpha }\right) \right) &=&P\left( S\left( R_{\alpha
}\right) \right) \cup \left\{ \left( a,b\right) ,\left( a,c\right) ,\left(
e,c\right) \right\} .
\end{eqnarray*}%
Fifth, for $\alpha \in \left[ 29/47,30/47\right) ,$

\begin{eqnarray*}
R_{\alpha } &=&R_{27/47}\setminus \left\{ \left( c,d\right) \right\} , \\
P\left( S\left( R_{\alpha }\right) \right) &=&R_{\alpha }=\left\{ \left(
a,d\right) ,\left( b,d\right) ,\left( c,b\right) ,\left( e,d\right) \right\}
\\
P\left( T\left( R_{\alpha }\right) \right) &=&P\left( S\left( R_{\alpha
}\right) \right) \cup \left\{ \left( c,d\right) \right\} .
\end{eqnarray*}%
Sixth, for $\alpha \in \left[ 30/47,31/47\right) ,$%
\begin{eqnarray*}
R_{\alpha } &=&R_{29/47}\setminus \left\{ \left( a,d\right) ,\left(
c,b\right) \right\} , \\
P\left( S\left( R_{\alpha }\right) \right) &=&R_{\alpha }=\left\{ \left(
b,d\right) ,\left( e,d\right) \right\} =P\left( T\left( R_{\alpha }\right)
\right) .
\end{eqnarray*}%
Seventh, for $\alpha \in \left[ 31/47,34/47\right) ,$ 
\begin{eqnarray*}
R_{\alpha } &=&R_{30/47}\setminus \left\{ \left( e,d\right) \right\} , \\
P\left( S\left( R_{\alpha }\right) \right) &=&R_{\alpha }=\left\{ \left(
b,d\right) \right\} =P\left( T\left( R_{\alpha }\right) \right) .
\end{eqnarray*}%
Eighth, for $\alpha \in \left[ 34/47,1\right) ,$%
\begin{equation*}
R_{\alpha }=S\left( R_{\alpha }\right) =\emptyset ,\text{ }\alpha \in \left[
34/47,1\right) .
\end{equation*}%
Among them, we explain the case where $\alpha \in \left[ 27/47,29/47\right) $%
. In this case, $R_{\alpha }$ is represented in \textbf{Figure 2}; that is,
a thick solid arrow from $j$ to $k$ exists if $\left( j,k\right) \in
R_{\alpha }$ or equivalently if 29 or more individuals prefer $j$ to $k$.
Since there is a P-cycle of $b\rightarrow d\rightarrow c\rightarrow b,$
(because $\left( b,d\right) ,\left( d,c\right) ,\left( c,b\right) \in
R_{\alpha }$)$,$ $\left( d,b\right) ,\left( c,d\right) ,\left( b,c\right)
\in S\left( R_{\alpha }\right) ,$ which are represented by the thin solid
arrows in Figure 2. Moreover, since $\left( a,d\right) ,\left( d,c\right)
\in R_{\alpha }$, $\left( a,c\right) \in T\left( R_{\alpha }\right)
\setminus S\left( R_{\alpha }\right) $. Similarly, $\left( a,b\right)
,\left( a,c\right) ,\left( e,b\right) ,\left( e,c\right) \in T\left(
R_{\alpha }\right) \setminus S\left( R_{\alpha }\right) $, which are
represented by the dot lines in Figure 2. Therefore, in this case, $S\left(
R_{\alpha }\right) \subsetneq T\left( R_{\alpha }\right) $.

\begin{figure}[tbph]
\begin{center}
\includegraphics[width=110mm]{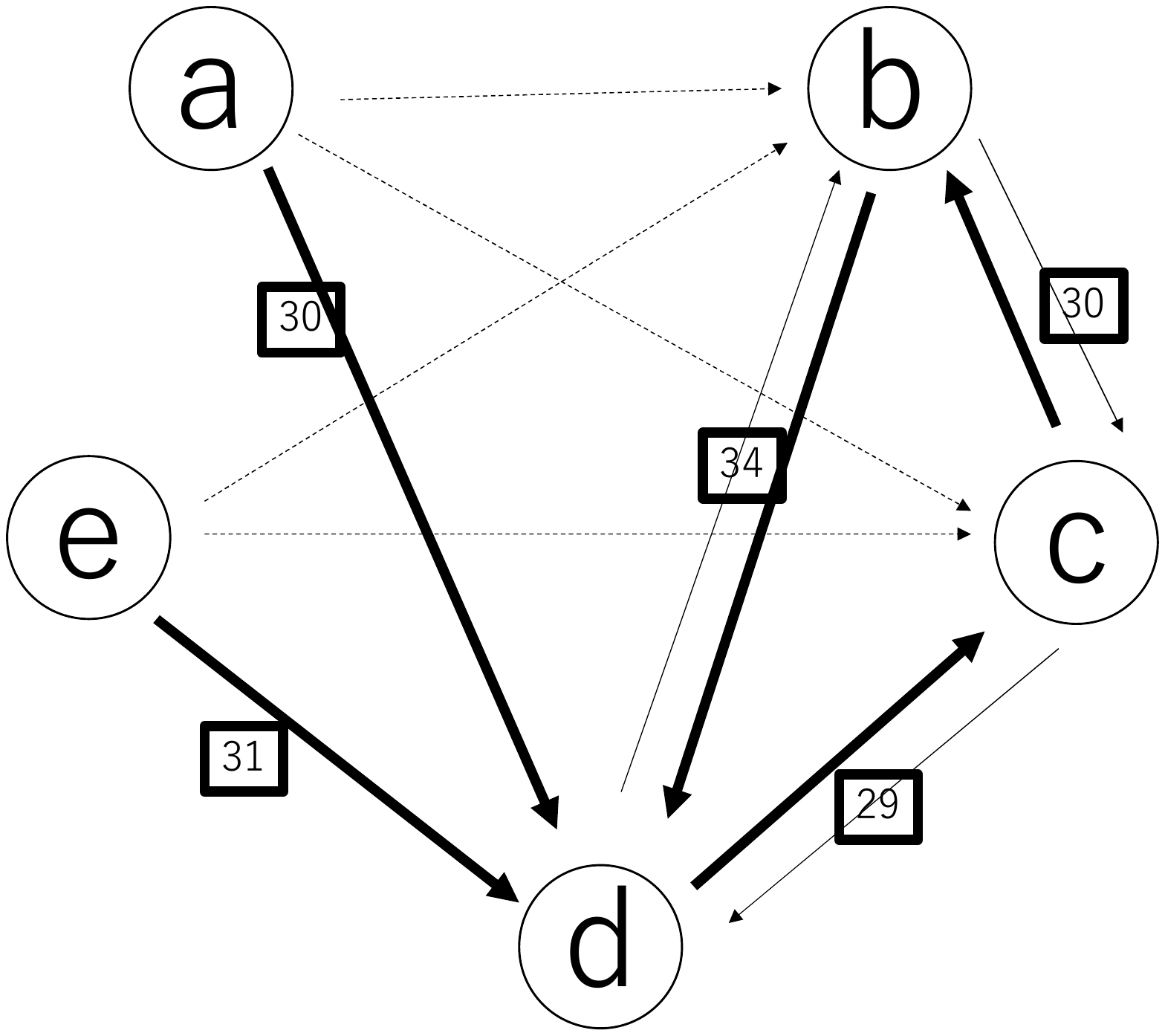}
\end{center}
\caption{$R_{\protect\alpha}$ of this example}
\end{figure}

Thus,%
\begin{eqnarray*}
\bigcup_{\alpha \in \left[ 1/2,1\right) }P\left( S\left( R_{\alpha }\right)
\right) &=&\left\{ \left( a,d\right) ,\left( b,d\right) ,\left( c,b\right)
,\left( e,b\right) ,\left( e,d\right) \right\} , \\
\bigcup_{\alpha \in \left[ 1/2,1\right) }P\left( T\left( R_{\alpha }\right)
\right) &=&\left( \bigcup_{\alpha \in \left[ 1/2,1\right) }P\left( S\left(
R_{\alpha }\right) \right) \right) \bigcup \\
&&\left\{ \left( a,b\right) ,\left( a,c\right) ,\left( c,d\right) ,\left(
e,a\right) ,\left( e,c\right) \right\} .
\end{eqnarray*}%
Then, we have 
\begin{eqnarray*}
\bigcap\limits_{\alpha \in \left[ 1/2,1\right) }\mathcal{R}\left( S\left(
R_{\alpha }\right) \right) &=&\mathcal{R}\left( \bigcup_{\alpha \in \left[
1/2,1\right) }P\left( S\left( R_{\alpha }\right) \right) \right) =\left\{
R_{1}^{\ast },\cdots ,R_{8}^{\ast }\right\} , \\
\bigcap\limits_{\alpha \in \left[ 1/2,1\right) }\mathcal{R}\left( T\left(
R_{\alpha }\right) \right) &=&\mathcal{R}\left( \bigcup_{\alpha \in \left[
1/2,1\right) }P\left( T\left( R_{\alpha }\right) \right) \right) =\left\{
R_{6}^{\ast }\right\} ,
\end{eqnarray*}%
where 
\begin{eqnarray*}
R_{1}^{\ast } &:&a,c,e,b,d,\text{ }R_{2}^{\ast }:a,e,c,b,d, \\
R_{3}^{\ast } &:&c,a,e,b,d,\text{ }R_{4}^{\ast }:c,e,a,b,d, \\
R_{5}^{\ast } &:&c,e,b,a,d,\text{ }R_{6}^{\ast }:e,a,c,b,d, \\
R_{7}^{\ast } &:&e,c,a,b,d,\text{ }R_{8}^{\ast }:e,c,b,a,d.
\end{eqnarray*}%
Therefore, the result of the Schulze method is $R_{6}^{\ast }$. On the other
hand, the result of the ranked pairs method is $R^{\pi }=R_{5}^{\ast }$ for
any $\pi \in \Pi $. Thus, the results of the two methods differ in this
example.

Next, we focus on other important voting methods to determine a collective
linear order. First, we consider the Kemeny-Young method. By this method, in
the example above, we have linear order $R^{KY}:$ $e,b,a,d,c.$ In this case, 
\begin{equation*}
R^{KY}\notin \bigcap\limits_{\alpha \in \left[ 1/2,1\right) }\mathcal{R}%
\left( S\left( R_{\alpha }\right) \right) ,
\end{equation*}%
because if $\alpha \in \left[ 29/47,30/47\right) ,$ then $\left( c,b\right)
\in P\left( S\left( R_{\alpha }\right) \right) =R_{\alpha }$.

Second, we consider the Borda method.\footnote{%
Since there are Condorcet cycles in $R_{1/2}$, the result of the Black
method introduced by Black (1958) is the same as that of the Borda rule.} By
this method, in the example above, we have linear order $R^{B}:$ $e,a,b,c,d$%
. Then, we also have%
\begin{equation*}
R^{B}\notin \bigcap\limits_{\alpha \in \left[ 1/2,1\right) }\mathcal{R}%
\left( S\left( R_{\alpha }\right) \right) ,
\end{equation*}%
because if $\alpha \in \left[ 29/47,30/47\right) ,$ then $\left( c,b\right)
\in P\left( S\left( R_{\alpha }\right) \right) =R_{\alpha }$.

\end{document}